\documentclass[aps,prx,twocolumn,english,superscriptaddress]{revtex4-2}

\usepackage{natbib}
\usepackage{amsfonts, amsmath, amssymb, mathrsfs, mathtools, amsthm, bbm, bm}
\usepackage{blkarray}
\usepackage{braket}
\usepackage{float}
\usepackage{afterpage}
\usepackage[version=4]{mhchem}
\usepackage[labelfont=bf, font=small]{caption}
\usepackage[font=footnotesize]{subcaption}
\usepackage{tikz}
\usetikzlibrary{decorations.pathreplacing} 
\usetikzlibrary{arrows}
\usepackage{quantikz}
\usepackage{booktabs, tabularx}

\usepackage{pifont}
\newcommand{\cmark}{\ding{51}}%
\newcommand{\xmark}{\ding{55}}%

\usepackage{xcolor,colortbl}
\newcommand{\na}{\cellcolor{black}[10pt]}  
\newcommand\Pad[1]{\hspace{1pt}#1\hspace{1pt}}

\DeclareMathOperator{\Tr}{Tr}
\DeclareMathOperator{\Var}{Var}

\newcommand{\E}[1]{\ensuremath{\mathcal{E}_{\mathrm{#1}}}}
\newcommand{\stat}[2]{\ensuremath{\mathrm{#1}(\E{#2})}}

\newcolumntype{Y}{>{\raggedleft\arraybackslash}X}

\usepackage[bookmarks=false]{hyperref}

\begin{document}
\title{Benchmarking Noisy Intermediate Scale Quantum Error Mitigation Strategies for Ground State Preparation of the \ce{HCl} Molecule}

\author{Tim Weaving}
\affiliation{Centre for Computational Science, Department of Chemistry, University College London, WC1H 0AJ, United Kingdom}
\author{Alexis Ralli}
\affiliation{Centre for Computational Science, Department of Chemistry, University College London, WC1H 0AJ, United Kingdom}
\author{William M. Kirby}
\affiliation{Department of Physics and Astronomy, Tufts University, Medford, MA 02155, USA}
\author{Peter J. Love}
\affiliation{Department of Physics and Astronomy, Tufts University, Medford, MA 02155, USA}
\affiliation{Computational Science Initiative, Brookhaven National Laboratory, Upton, NY 11973, USA}
\author{Sauro Succi}
\affiliation{Center for Life Nano-Neuro Science @ La Sapienza, Italian Institute of Technology, 00161 Roma, Italy}
\affiliation{Department of Mechanical Engineering, University College London, WC1E 7JE, United Kingdom}
\affiliation{Department of Physics, Harvard University, Cambridge, MA 02138, USA}
\author{Peter V. Coveney}
\affiliation{Centre for Computational Science, Department of Chemistry, University College London, WC1H 0AJ, United Kingdom}
\affiliation{Advanced Research Computing Centre, University College London, WC1H 0AJ, United Kingdom}
\affiliation{Informatics Institute, University of Amsterdam, Amsterdam, 1098 XH, Netherlands}

\date{\today}

\begin{abstract}
Due to numerous limitations including restrictive qubit topologies, short coherence times and prohibitively high noise floors, few quantum chemistry experiments performed on existing noisy intermediate-scale quantum hardware have achieved the high bar of chemical precision, namely energy errors to within 1.6 mHa of full configuration interaction. To have any hope of doing so, we must layer contemporary resource reduction techniques with best-in-class error mitigation methods; in particular, we combine the techniques of qubit tapering and the contextual subspace variational quantum eigensolver with several error mitigation strategies comprised of measurement-error mitigation, symmetry verification, zero-noise extrapolation and dual-state purification. We benchmark these strategies across a suite of eight 27-qubit IBM Falcon series quantum processors, taking preparation of the \ce{HCl} molecule's ground state as our testbed.
\end{abstract}

\maketitle

\section{Introduction}\label{sec:intro}

We find ourselves in the era of noisy intermediate-scale quantum (NISQ) computation, which is characterized by various obstacles including restrictive qubit topologies, short coherence times and imperfect quantum gates; these factors compound to limit what is achievable using existing or near-term quantum devices. The development of quantum error mitigation (QEM) techniques has therefore been a necessary pursuit, aiming to extract usable data from the raw output of NISQ machines.

A plethora of techniques have been proposed that exploit various properties of noise in NISQ devices. Some evaluate collections of Clifford circuits (which are classically efficient to simulate) either to mitigate against measurement errors \cite{bravyi2021mitigating, nation2021scalable} or for learning-based approaches that aim to characterize the noise model \cite{strikis2021learning, czarnik2021error, lowe2021unified}. Others average over the effect of noise by recompiling circuits at random \cite{temme2017error, endo2018practical, mari2021extending} or make predictions informed by the character of the noise and its behaviour under amplification \cite{li2017efficient, temme2017error, endo2018practical, kandala2019error, giurgica2020digital, he2020zero, mari2021extending}. There are also techniques that use problem-specific properties (e.g. known symmetries) to identify and discard invalid outcomes \cite{bonet2018low, mcardle2019error, cai2021quantum} and purification-based approaches that promote some pure component of the noisy states prepared in hardware \cite{huggins2021virtual, o2021error, cai2021resource, czarnik2021qubit, huo2022dual}. We refer the reader to the works of Endo \textit{et al.} \cite{endo2021hybrid} and Cai \textit{et al.} \cite{cai2022quantum} for a comprehensive review of the QEM literature and to Resch \& Karpuzcu \cite{resch2021benchmarking} for an exposition of noise sources in quantum computation.

While QEM has permitted some degree of success in obtaining usable results from NISQ computers, a number of works have cautioned that QEM may be restricted by some fundamentals limits \cite{takagi2022fundamental, takagi2022universal, quek2022exponentially}. With this in mind, it is not clear whether `quantum advantage' will be feasible using QEM alone and we may still require partially error corrected machines for this to be realised in practice.

\begin{figure}[b]
    \centering
    \setlength\tabcolsep{0pt}
      \begin{tabular}{ | r | c | c | c | c | c | } \hline
        \na & \Pad{MEM} & SV & ZNE & DSP & TP \\
        \hline
        \Pad{MEM} & \na & \cmark & \cmark & \cmark & \xmark \\ \hline
        \Pad{SV} & \cmark & \na & \cmark & \xmark & \xmark \\ \hline
        \Pad{ZNE} & \cmark & \cmark & \na& \cmark & \xmark \\ \hline
        \Pad{DSP} & \cmark & \xmark & \cmark & \na & \cmark \\ \hline
        \Pad{TP} & \xmark & \xmark & \xmark & \cmark & \na \\
        \hline
      \end{tabular}
    \caption{Compatibility matrix of the error-mitigation techniques investigated in this work. Note that tomography purification is compatible with each of these techniques in principle, however it is not in general scalable due to the exponential number of bases one must measure to reconstruct the density matrix via full state tomography. We claim compatibility with dual-state purification since we need only apply it to a single ancilla qubit.}
    \label{fig:compatibility_matrix}
\end{figure}

In this work we place an emphasis on \textit{scalable} quantum error mitigation techniques for the NISQ era. As such, we benchmark the following:
\begin{enumerate}
    \item Measurement-error mitigation (MEM) - \ref{sec:MEM}
    \item Non-$\mathbb{Z}_2$ symmetry verification (SV) - \ref{sec:SV}
    \item Zero-noise extrapolation (ZNE) - \ref{sec:ZNE}
    \item Dual-state purification (DSP) - \ref{sec:DSP}
    \item Tomography purification (TP) applied to DSP
\end{enumerate}
including every possible combination given by the compatibility matrix in Figure \ref{fig:compatibility_matrix}. For a fixed shot budget we intend to identify which combined strategy is most effective in mitigating errors, executed across a suite of IBM quantum hardware.

The problem we take as a testbed for this QEM benchmark is preparation of the $\ce{HCl}$ molecule ground state, with the ultimate goal of measuring the corresponding energy to chemical precision (errors within 1.6 mHa of full configuration interaction). Of the numerous quantum chemistry experiments performed on NISQ hardware to date \cite{peruzzo2014variational,Shen2017,OMalleyBabbush2016,Santagati2018a,Kandala2017,Colless2018,hempel2018quantum,Kandala2019,nam2020ground,McCaskey2019,Smart2019,arute2020hartree,Gao2021a,Kawashima2021,Rice2021,eddins2022doubling,Motta2022,Yamamoto2022,Kirsopp2022,Khan2022,OBrien2022,Zhao2022,kiss2022quantum}, only a select few have achieved this threshold; of those that have, most consist of hydrogen chains of varying size. 

\section{The Hardware}\label{sec:hardware}

The IBM Quantum hardware is equipped with the universal gate set $\{\mathrm{CNOT}, R_z, X, \sqrt{X}\}$ and, at the time of writing, eight $27$-qubit Falcon series quantum processors were available to us. From the point of view of gate errors and coherence these devices are the most reliable available through IBM Quantum at present, with the greatest Quantum Volumes (QV) \cite{cross2019validating, resch2021benchmarking}; in Table \ref{hardware_specs} we provide a snapshot of the hardware specification at the point of execution of our Qiskit Runtime programs.

\begin{figure}[h!]
    \centering
    \begin{quantikz}[row sep=0.1cm, column sep = 0.3cm]
        \lstick[wires=7]{$N$-qubits}
                  & \gate{H} & \ctrl{1} & \qw      & \qw       & \qw      & \qw      & \qw \\
        \lstick{} & \qw      & \targ{}  & \ctrl{1} & \qw       & \qw      & \qw      & \qw \\
        \lstick{} & \qw      & \qw      & \targ{}  & \ctrl{1}  & \qw      & \qw      & \qw \\
                  &          &          &          & \vdots    &          &          &     \\   
        \lstick{} & \qw      & \qw      & \qw      & \targ{}   & \ctrl{1} & \qw      & \qw \\
        \lstick{} & \qw      & \qw      & \qw      & \qw       & \targ{}  & \ctrl{1} & \qw \\
        \lstick{} & \qw      & \qw      & \qw      & \qw       & \qw      & \targ{}  & \qw  
    \end{quantikz}
    \caption{The $N$-qubit GHZ circuit, consisting of one Hadamard and $N-1$ CNOT gates.}
    \label{fig:GHZ_circuit}
\end{figure}
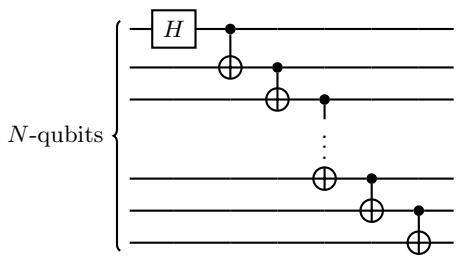

One way we may assess the quality of these devices is to evaluate quantum state fidelities for increasing numbers of qubits. Namely, we shall prepare the $N$-qubit Greenberger–Horne–Zeilinger (GHZ) state
\begin{equation}
    \ket{\psi_N} = \big(\ket{0}^{\otimes N} + \ket{1}^{\otimes N}\big) / \sqrt{2}
\end{equation}
via the circuit given in Figure \ref{fig:GHZ_circuit} and determine the fidelity
\begin{equation}
\begin{aligned}
    f(N) 
    ={} & | \braket{\psi^{\mathrm{true}}_N | \psi_N^{\mathrm{noisy}}} |^2 \\
    ={} & \frac{1}{2} (\sqrt{p_{\bm{0}}} + \sqrt{p_{\bm{1}}})^2
\end{aligned}
\end{equation}
where $\psi_N^{\mathrm{noisy}}$ is the noisy state prepared on the hardware and $p_{\bm{0}}, p_{\bm{1}}$ are the probabilities with which we obtain the all $\bm{0}$ or $\bm{1}$ state, respectively.

\begin{figure}[t!]
    \centering
    \includegraphics[width=\linewidth]{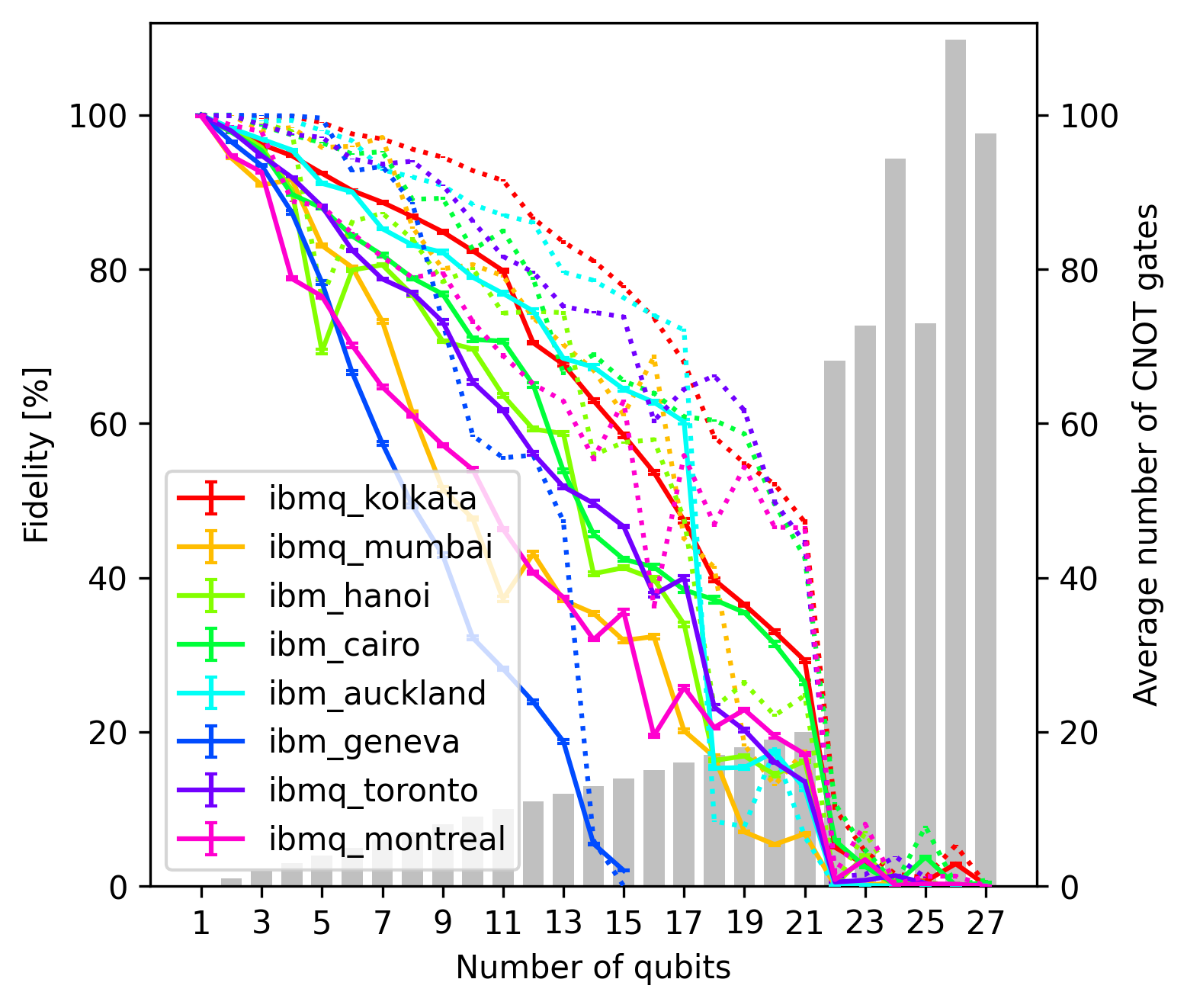}
    \caption{Decay in quantum state fidelity against number of qubits for GHZ preparation; dotted lines indicate the measurement-error mitigated result. We could not utilize more than 15-qubits on \textit{ibm\_geneva} due to a faulty qubit. The grey bars indicate the average number of CNOT gates required to prepare the relevant $N$-qubit GHZ state, with differences arising between chips due to the transpiler.}
    \label{fig:GHZ_benchmark}
\end{figure}

\begin{figure}[b!]
    \centering
    \begin{tikzpicture}[shorten >=1pt, auto, node distance=10mm,
     every node/.style={circle,thick,draw,minimum size=8mm}
     ]
    \node[left color=green, right color=orange] (A)              {$q_0$};    \node[left color=green, right color=orange] (B) [right of=A] {$q_1$};    \node[left color=green, right color=orange] (C) [below of=B] {$q_4$};
    \node[fill=green] (D) [below of=C] {$q_7$};    \node (E) [below of=D] {$q_{10}$}; \node (F) [below of=E] {$q_{12}$};
    \node (G) [below of=F] {$q_{15}$}; \node (H) [below of=G] {$q_{18}$}; \node (I) [below of=H] {$q_{21}$};
    \node (J) [below of=I] {$q_{23}$};
    \node[left color=green, right color=orange] (K) [right of=B] {$q_2$};
    \node[fill=orange] (L) [right of=K] {$q_3$};    \node[fill=yellow] (M) [below of=L] {$q_5$};    \node[left color=pink, right color=yellow] (N) [below of=M] {$q_8$};
    \node[left color=pink, right color=yellow] (O) [below of=N] {$q_{11}$}; \node[left color=pink, right color=yellow] (P) [below of=O] {$q_{14}$}; \node[fill=pink] (Q) [below of=P] {$q_{16}$};
    \node (R) [below of=Q] {$q_{19}$}; \node (S) [below of=R] {$q_{22}$}; \node (T) [below of=S] {$q_{25}$};
    \node (U) [right of=T] {$q_{26}$};
    \node (V) [left of=T] {$q_{24}$}; \node[fill=pink] (W) [right of=F] {$q_{13}$}; \node[fill=yellow] (X) [right of=N] {$q_9$};
    \node (Y) [right of=R] {$q_{20}$}; \node (Z) [left of=D] {$q_6$}; \node (ZA)[left of=H] {$q_{17}$};
    \draw (A) edge (B); \draw (B) edge (C); \draw (C) edge (D); \draw (D) edge (E); \draw (E) edge (F); 
    \draw (F) edge (G); \draw (G) edge (H); \draw (H) edge (I); \draw (I) edge (J); \draw (F) edge (G);
    \draw (B) edge (K); \draw (K) edge (L); \draw (L) edge (M); \draw (M) edge (N); \draw (N) edge (O);
    \draw (O) edge (P); \draw (P) edge (Q); \draw (Q) edge (R); \draw (R) edge (S); \draw (S) edge (T);
    \draw (T) edge (U); \draw (T) edge (V); \draw (V) edge (J); \draw (F) edge (W); \draw (D) edge (Z);
    \draw (P) edge (W); \draw (N) edge (X); \draw (R) edge (Y); \draw (H) edge (ZA);
    \end{tikzpicture}
    
    \caption{The IBM \textit{Falcon} series 27-qubit chip `heavy-hex' topology. For our quantum simulations we identified optimal qubit clusters by assigning scores based on gate and readout errors. For DSP we require 5-qubit clusters of the form given in Figure \ref{fig:dsp_cluster} to facilitate every possible readout configuration; we have highlighted the specific clusters used, as detailed in Table \ref{hardware_specs}}
    \label{fig:kolkata_topology}
\end{figure}
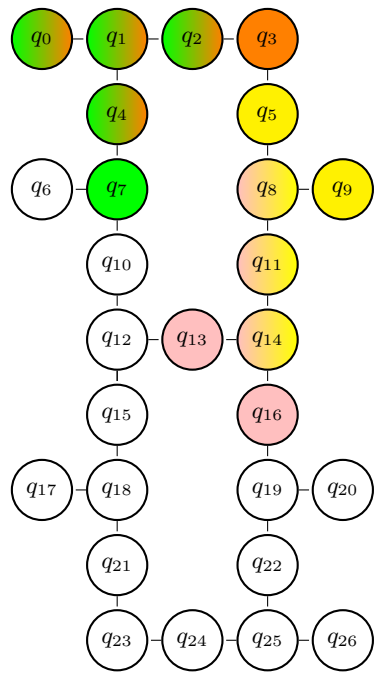

\begin{table*}
\centering
\caption{Breakdown of quantum hardware specification restricted to the chosen qubit cluster at the point of executing the Qiskit Runtime programs. We provide the Quantum Volume (QV), chosen 5-qubit cluster, T1/T2 times and gate duration/error for entangling (CNOT), local ($R_z, X, \sqrt{X}$) and readout operations.}
\label{hardware_specs}
\begin{tabular}{lrlrrrrr}
\toprule
            &      &                     & \multicolumn{2}{c}{Coherence}            &         \multicolumn{3}{c}{Gate Specification} \\
            &   QV &   Chosen 5q Cluster & Type          &            Time [$\mu$S] &         Type &        Time [nS] & Error $\times 10^3$ \\
\midrule
ibmq\_montreal 
            &  128 &   \{0, 1, 2, 3, 4\} &  \textbf{T1}: &     $ 140.92 \pm  16.77$ &  \textbf{Entangling}: &   $ 471.11 \pm  78.69$ &         $ 7.85 \pm  1.06$ \\
            &      &                     &  \textbf{T2}: &      $ 82.16 \pm  39.10$ &       \textbf{Local}: &     $ 35.56 \pm  0.00$ &         $ 0.22 \pm  0.03$ \\
            &      &                     &               &                          &     \textbf{Readout}: &   $ 5201.78 \pm  0.00$ &        $ 14.08 \pm  2.52$ \\
ibmq\_kolkata 
            &  128 & \{16, 19, 22, 25, 20\} &  \textbf{T1}: &     $ 150.92 \pm  16.80$ &  \textbf{Entangling}: &  $ 348.44 \pm  177.77$ &         $ 5.14 \pm  0.65$ \\
            &      &                     &  \textbf{T2}: &     $ 135.59 \pm  66.36$ &       \textbf{Local}: &     $ 35.56 \pm  0.00$ &         $ 0.17 \pm  0.04$ \\
            &      &                     &               &                          &     \textbf{Readout}: &    $ 675.56 \pm  0.00$ &        $ 10.68 \pm  1.99$ \\
ibmq\_mumbai 
            &  128 &   \{0, 1, 2, 3, 4\} &  \textbf{T1}: &     $ 129.80 \pm  28.12$ &  \textbf{Entangling}: &  $ 556.44 \pm  136.77$ &         $ 8.63 \pm  2.20$ \\
            &      &                     &  \textbf{T2}: &     $ 104.20 \pm  69.62$ &       \textbf{Local}: &     $ 35.56 \pm  0.00$ &         $ 0.31 \pm  0.17$ \\
            &      &                     &               &                          &     \textbf{Readout}: &   $ 3552.00 \pm  0.00$ &        $ 18.24 \pm  0.91$ \\
ibm\_hanoi 
            &   64 &       \{0, 1, 4, 7, 2\} &  \textbf{T1}: &     $ 135.29 \pm  54.61$ &  \textbf{Entangling}: &   $ 270.67 \pm  59.21$ &         $ 6.99 \pm  2.65$ \\
            &      &                     &  \textbf{T2}: &     $ 174.89 \pm  83.31$ &       \textbf{Local}: &     $ 32.00 \pm  0.00$ &         $ 0.21 \pm  0.10$ \\
            &      &                     &               &                          &     \textbf{Readout}: &    $ 817.78 \pm  0.00$ &         $ 8.94 \pm  1.59$ \\
ibm\_cairo 
            &   64 &   \{16, 14, 11, 8, 13\} &  \textbf{T1}: &      $ 95.59 \pm  37.85$ &  \textbf{Entangling}: &  $ 462.22 \pm  296.50$ &         $ 9.17 \pm  4.30$ \\
            &      &                     &  \textbf{T2}: &      $ 92.76 \pm  69.45$ &       \textbf{Local}: &     $ 24.89 \pm  0.00$ &         $ 0.22 \pm  0.06$ \\
            &      &                     &               &                          &     \textbf{Readout}: &    $ 732.44 \pm  0.00$ &       $ 20.86 \pm  11.74$ \\
ibm\_auckland 
            &   64 &   \{16, 14, 11, 8, 13\} &  \textbf{T1}: &     $ 162.99 \pm  73.58$ &  \textbf{Entangling}: &   $ 376.89 \pm  28.61$ &         $ 6.28 \pm  1.22$ \\
            &      &                     &  \textbf{T2}: &     $ 123.67 \pm  72.80$ &       \textbf{Local}: &     $ 35.56 \pm  0.00$ &         $ 0.23 \pm  0.03$ \\
            &      &                     &               &                          &     \textbf{Readout}: &    $ 757.33 \pm  0.00$ &         $ 8.30 \pm  1.55$ \\
ibmq\_toronto 
            &   32 &     \{9, 8, 11, 14, 5\} &  \textbf{T1}: &      $ 113.96 \pm  6.53$ &  \textbf{Entangling}: &   $ 382.22 \pm  61.35$ &         $ 7.96 \pm  0.83$ \\
            &      &                     &  \textbf{T2}: &     $ 171.38 \pm  19.39$ &       \textbf{Local}: &     $ 35.56 \pm  0.00$ &         $ 0.28 \pm  0.06$ \\
            &      &                     &               &                          &     \textbf{Readout}: &   $ 5962.67 \pm  0.00$ &        $ 12.04 \pm  4.30$ \\
ibm\_geneva &   32 &     \{9, 8, 11, 14, 5\} &  \textbf{T1}: &    $ 311.27 \pm  101.19$ &  \textbf{Entangling}: &  $ 586.67 \pm  110.22$ &         $ 3.97 \pm  0.31$ \\
            &      &                     &  \textbf{T2}: &    $ 300.56 \pm  127.31$ &       \textbf{Local}: &     $ 38.40 \pm  5.69$ &         $ 0.20 \pm  0.17$ \\
            &      &                     &               &                          &     \textbf{Readout}: &   $ 1600.00 \pm  0.00$ &       $ 30.26 \pm  13.08$ \\
\bottomrule
\end{tabular}
\end{table*}

In Figure \ref{fig:GHZ_benchmark} we observe a decay in fidelity as more qubits are included in the GHZ state preparation procedure, with a sharp drop to near-zero fidelity at $N=22$. This is due to the longest connected path of qubits being of length $21$, given the chip topology of Figure \ref{fig:kolkata_topology}; beyond this point we incur expensive SWAP operations that rapidly consume the remaining fidelity, indicated by the dramatic jump in number of CNOT gates from 22-qubit onwards. We also include the effect of measurement-error mitigation on the fidelity and note that we are able to recover approximately $10-20\%$ fidelity in most cases.

\section{Qubit Reduction Techniques}\label{sec:QR_techniques}

Taken in the minimal STO-3G basis, the full \ce{HCl} problem consists of 20 qubits and therefore direct treatment is not yet feasible on current quantum computers. In order for the hardware to accommodate our problem we layered the qubit reduction techniques of tapering \cite{bravyi2017tapering, setia2020reducing} (Section \ref{sec:tapering}) and contextual subspace \cite{kirby2021contextual, weaving2023stabilizer, ralli2023unitary} (Section \ref{sec:contextual}) to yield a dramatically condensed 3-qubit Hamiltonian
\begin{equation}
    H = \sum_{i} h_i P_i
\end{equation}
where we provide the explicit coefficients $h_i \in \mathbb{R}$ and Pauli terms $P_i = q_0^{(i)} \otimes q_1^{(i)} \otimes q_2^{(i)}$ in Table \ref{tab:3q_hamiltonian}. The exact ground state energy of this Hamiltonian lies within $0.837$ mHa of the full configuration interaction (FCI) energy ($-455.157$Ha, calculated using PySCF \cite{sun2018pyscf}); this is nearly half what is generally considered chemical precision (1.6 mHa), although we stress that, due to the minimal basis set used here, one should not expect agreement with experimentally-obtained energy values. Subtracting the relatively large identity term leaves a target energy of $-2.066$ Ha; with respect to chemical precision, this represents a challenging $0.077\%$ error ratio that we aim to capture via QEM.

Due to incompatibility with some of the error-mitigation techniques investigated here, we do not implement any measurement reduction strategies such as (qubit-wise) commuting decompositions or unitary partitioning \cite{izmaylov2019unitary, ralli2021implementation}. Instead, each Hamiltonian term is treated independently so there is zero covariance between expectation value estimates and the overall variance is therefore obtained as
\begin{equation}
   \Var(H) = \sum_{i} h_i^2 \cdot \Var(P_i);
\end{equation}
the statistical analysis is conducted with a bootstrapping of the raw quantum measurement data.

\subsection{Qubit Tapering}\label{sec:tapering}

Tapering allows one to map Hamiltonian $\mathbb{Z}_2$ symmetries onto distinct qubits and consequently project over them, thus reducing the effective dimension of the problem. This works by identifying an independent set of Pauli operators $\mathcal{S} \subset \mathcal{P}_N$ such that $[S, T] = 0 \;\forall\, S \in \mathcal{S}, T \in \mathcal{T}$, which we refer to as \textit{symmetry generators} and can be identified efficiently using the \textit{Symmer} Python package \cite{symmer2022}. Assuming the elements of $\mathcal{S}$ commute amongst themselves (if not, select the largest commuting subset within) one may perform a Clifford rotation mapping each symmetry to a distinct qubit position and consequently project onto the corresponding stabilizer subspace; under this procedure it is possible to remove $|\mathcal{S}|$ qubits from the Hamiltonian while remaining isospectral.

Since this is a fermionic system we are guaranteed a reduction of at least two qubits arising from the preservation of spin up/down parities; under the Jordan-Wigner mapping \cite{jordan1993paulische} these manifest as $S_\mathrm{up/down} = Z^{\otimes \mathcal{I}_{\mathrm{up/down}}}$ where the sets $\mathcal{I}_{\mathrm{up}}, \mathcal{I}_{\mathrm{down}}$ index qubit positions encoding up ($\alpha$), down ($\beta$) electron spin orbitals, respectively. These spin parity operators are still $\mathbb{Z}_2$ symmetries (i.e. single-Paulis terms) under the Bravyi-Kitaev mapping \cite{bravyi2002fermionic}, however their closed form is less convenient since individual qubits do not represent distinct spin-orbitals. For our particular formulation of the $20$-qubit \ce{HCl} system with even (odd) indices encoding spin up (down) electrons we have
\begin{equation}\label{up_down_parity}
\begin{aligned}
    S_{\mathrm{up}} ={} & ZIZIZIZIZIZIZIZIZIZI, \\
    S_{\mathrm{down}} ={} & IZIZIZIZIZIZIZIZIZIZ.
\end{aligned}
\end{equation}
We also identified two additional $\mathbb{Z}_2$ symmetries
\begin{equation}
\begin{aligned}
    S_{\sigma_h} ={} & IIIIIIIIZZIIIIIIZZII, \\
    S_{C_2} ={} & IIIIIIZZZZIIIIZZZZII,
\end{aligned}
\end{equation}
that arise from the abelian subgroup $C_{2v}$ of the non-abelian point group $C_{\infty v}$ (to which all heteronuclear diatomic molecules belong) generated by reflections along the molecular plane ($\sigma_h$ symmetry) and rotations through an angle of $180^\circ$ ($C_2$ symmetry). In all, with the symmetry generating set $\mathcal{S} = \{S_{\sigma_h}, S_{C_2}, S_{\mathrm{up}}, S_{\mathrm{down}}\}$, qubit tapering permits a reduction of 20 to 16 qubits while exactly preserving the energy spectrum. 


\subsection{Contextual Subspace}\label{sec:contextual}

Whereas tapering exploits physical symmetries of the Hamiltonian to remove redundant qubits, it is possible to achieve further reductions by imposing pseudo-symmetries on the system. This is the contextual subspace approach \cite{kirby2021contextual, weaving2023stabilizer, ralli2023unitary} in which we partition the Hamiltonian into noncontextual and contextual components; the former may be mapped onto a classical optimization problem whereas the latter yields quantum corrections obtained via some eigenvalue-finding algorithm (VQE, QPE etc.). The qubit reduction is effected by enforcing noncontextual symmetries on the contextual Hamiltonian, thus ensuring any quantum corrections are consistent with the noncontextual ground state configuration.

The choice over which noncontextual symmetries to enforce is highly non-trivial. Here, we select stabilizers that preserve commutativity with the most dominant coupled-cluster amplitudes, thus maximising variational flexibility in the contextual subspace. Using this heuristic, we are able to project onto a 3-qubit contextual subspace that permits chemical precision. In Table \ref{tab:3q_hamiltonian} we provide explicit details of the corresponding Hamiltonian, whose ground state energy has absolute error $0.837$mHa with respect to the FCI energy.

This dramatic reduction in qubit resource is likely due to CCSD being near exact (we obtained an error of $3.403 \times 10^{-8}$ Ha with respect to FCI, five orders of magnitude below chemical precision), as there are just two unoccupied spin-orbitals in the minimal STO-3G basis set and therefore excitations above doubles are not possible.


\begin{table}[h]
    \centering
\begin{tabular}{lcccr|lcccr}
\toprule
Index & $q_0$ & $q_1$ & $q_2$ &  Coefficient & Index & $q_0$ & $q_1$ & $q_2$ &  Coefficient \\
\midrule
0  &   I &   I &   I &  -453.090742 & 17 &   Y &   Y &   X &     0.035219 \\
1  &   I &   Z &   Z &     0.846721 & 18 &   I &   I &   X &    -0.015458 \\
2  &   Z &   I &   Z &     0.846721 & 19 &   I &   Z &   X &     0.015458 \\
3  &   I &   Z &   I &     0.620754 & 20 &   Z &   I &   X &     0.015458 \\
4  &   Z &   I &   I &     0.620754 & 21 &   Z &   Z &   X &    -0.015458 \\
5  &   I &   I &   Z &     0.393828 & 22 &   I &   X &   X &    -0.009644 \\
6  &   Z &   Z &   I &     0.258369 & 23 &   I &   Y &   Y &    -0.009644 \\
7  &   Z &   Z &   Z &     0.238049 & 24 &   Z &   X &   X &     0.009644 \\
8  &   X &   Z &   I &    -0.061959 & 25 &   Z &   Y &   Y &     0.009644 \\
9  &   Z &   X &   I &     0.061959 & 26 &   X &   I &   X &     0.009644 \\
10 &   Z &   X &   Z &    -0.061959 & 27 &   X &   Z &   X &    -0.009644 \\
11 &   X &   Z &   Z &     0.061959 & 28 &   Y &   I &   Y &     0.009644 \\
12 &   Y &   Y &   I &    -0.055599 & 29 &   Y &   Z &   Y &    -0.009644 \\
13 &   Y &   Y &   Z &     0.055599 & 30 &   I &   X &   I &     0.004504 \\
14 &   X &   X &   X &    -0.035219 & 31 &   I &   X &   Z &    -0.004504 \\
15 &   X &   Y &   Y &    -0.035219 & 32 &   X &   I &   I &    -0.004504 \\
16 &   Y &   X &   Y &    -0.035219 & 33 &   X &   I &   Z &     0.004504 \\
\bottomrule
\end{tabular}
\caption{The 3-qubit contextual subspace \ce{HCl} Hamiltonian, terms ordered by coefficient magnitude, we take as a testbed for the error-mitigation strategies investigated in this work.}
\label{tab:3q_hamiltonian}
\end{table}

\section{Error Mitigation}\label{sec:QEM}

In this section we review the technical aspects of each quantum error mitigation (QEM) technique investigated through our benchmark, the results of which are later discussed in Section \ref{sec:results}. 

\subsection{Estimators}

The language we shall use to describe our QEM techniques is that of \textit{estimators}. Suppose that we are interested in some observable $O$ (a Hermitian operator, i.e. $O^\dag = O$) and have access to a general quantum state $\rho$; we wish to estimate the quantity $\Tr{(\rho O)}$, but may only probe the state via some finite sample of quantum measurements $\mathcal{M} = \{m_i\}_{i=1}^M$ where $m_i \in \mathbb{Z}_2^N$. The way in which we collect and subsequently combine our sample to approximate the desired observable property defines an estimator $\mathcal{E}: \mathcal{M} \rightarrow \mathbb{R}$; the goal of QEM is to construct effective estimators that are capable of suppressing errors and extracting some usable data from the noise.

For example, we may define a na\"ive estimator for the expectation value of a Pauli operator  $P \in \mathcal{P}_N$. Given a pure quantum state $\ket{\psi}$, we may sample from the quantum device in a compatible basis (i.e. one that commutes with $P$) and obtain eigenstates $\ket{m_i}$ such that $P \ket{m_i} = m_i \ket{m_i}$ where $m_i = \pm1$ to estimate the expectation value $\braket{P}_\psi \coloneqq \bra{\psi} P \ket{\psi}$. The raw estimator is
\begin{equation}
    \E{RAW}^P(\mathcal{M}) = \frac{1}{M} \sum_{i=1}^M m_i \rightarrow \braket{P}_\psi\;\; (M \rightarrow \infty).
\end{equation}
Since any Hermitian operator may be decomposed as $O = \sum_P o_P P$ with $o_P \in \mathbb{R}$, this allows us to extend our estimator to the full observable by linearity
\begin{equation}
    \E{RAW} = \sum_P o_P \E{RAW}^P,
\end{equation}
which shall form a baseline for our QEM benchmark. 

We shall use the following metrics to assess the efficacy of QEM techniques:
\begin{equation}
\begin{aligned}
    \stat{var}{} ={} & \mathbb{E}(\mathcal{E}^2) - \mathbb{E}(\mathcal{E})^2 \\
    \stat{bias}{} ={} & \mathbb{E}(\mathcal{E} - \braket{O}_\psi)
\end{aligned}
\end{equation}
and the related quantity
\begin{equation}
\begin{aligned}
    \stat{MSE}{} ={} & \mathbb{E}\Big((\mathcal{E}-\braket{O}_\psi)^2\Big) \\
    ={} & \mathrm{var}(\mathcal{E}) + \mathrm{bias}(\mathcal{E})^2,
\end{aligned}
\end{equation}
or mean squared error. Taking $O = H$ and $\ket{\psi}$ the ground state of $H$, our objective is to approximate $\mathbb{E}(\mathcal{E}) \approx \bra{\psi} H \ket{\psi} = \braket{H}_{\psi} = E_{\mathrm{FCI}}$. The goal of QEM is to reduce bias as far as possible (ideally within the threshold of chemical precision, i.e. $|\stat{bias}{QEM}|<1.6$ mHa) while aiming not to amplify variance severely.


Although it would be preferable to run multiple instances of each quantum simulation to evaluate $\mathbb{E}(\mathcal{E})$, this is not feasible given the length of time taken to produce each energy estimate. Instead, we rely on the statistical tool of \textit{bootstrapping}, introduced in further detail in Appendix \ref{sec:bootstrapping}, whereby we generate resampled data from the empirical measurement outcomes. 

\subsection{Measurement-Error Mitigation}\label{sec:MEM}



Measurement-error mitigation (MEM) aims to characterize the errors incurred during the readout phase of a quantum experiment \cite{bravyi2021mitigating}; it treats the state preparation itself as a black box and does not consider errors that occur prior to measurement. 

A naive, non-scalable, approach to MEM is to prepare-and-measure each of the $2^N$ basis states individually; given some $\ket{\bm{b}_i}$ with $\bm{b}_i \in \mathbb{Z}_2^{N}$ we perform measurements to obtain a noisy distribution of binary outcomes $\ket{\mu^{(i)}_{\mathrm{noisy}}} = \sum_j p_{i,j} \ket{\bm{b}_j}$ where $p_{i,j} = \bra{\bm{b}_i} A \ket{\bm{b}_j}$ denotes the probability of preparing the state $\ket{\bm{b}_i}$ and measuring $\ket{\bm{b}_j}$. The doubly stochastic matrix $A = \sum_{i,j} p_{i,j} \ket{\bm{b}_j} \bra{\bm{b}_i}$ is referred to as the assignment (or transition) matrix and lies at the core of this technique. 

Now, suppose we wish to implement a circuit with noiseless measurement output $\ket{\mu_{\mathrm{ideal}}} = \sum_{i} m_i \ket{\bm{b}_i}$; since $A \ket{\bm{b}_i} = \ket{\mu^{(i)}_{\mathrm{noisy}}}$, then by linearity we have 
\begin{equation}
    A \ket{\mu_{\mathrm{ideal}}} = \sum_i m_i \ket{\mu^{(i)}_{\mathrm{noisy}}} =: \ket{\mu_{\mathrm{noisy}}}.
\end{equation}
More realistically, what we will actually have access to is $\ket{\mu_{\mathrm{noisy}}}$, the output from some quantum experiment. Therefore, by inverting the assignment matrix we obtain a measurement-error mitigated distribution $\ket{\mu_{\mathrm{ideal}}} = A^{-1} \ket{\mu_{\mathrm{noisy}}}$.

In its current form, it will not be possible to construct the assignment matrix for large numbers of qubits. The `tensored' approach of Nation \textit{et al.} \cite{nation2021scalable} is designed to assess the \textit{qubitwise} measurement assignment error, namely evaluating the probability $p_k$ that qubit $k$ is eroneously flipped $\ket{0} \rightleftharpoons \ket{1}$. The single-qubit assignment matrix for this process is
\begin{equation}
    A^{(k)} = 
    \begin{blockarray}{ccc}
        & \ket{0} & \ket{1} \\
        \begin{block}{c[cc]}
            \bra{0} & 1-p_k & p_k \\
            \bra{1} & p_k & 1-p_k \\        
        \end{block}
\end{blockarray}
\end{equation}
and we subsequently reconstruct the full $N$-qubit assignment error probability by taking products over the relevant single-qubit transitions
\begin{equation}\label{sq_readout}
    A_{i,j} \approx \prod_{k=0}^{N-1} A^{(k)}_{(\bm{b}_i)_k, (\bm{b}_j)_k}.
\end{equation}
This expression makes some strong assumptions on the character of the readout errors, in particular that they are predominantly uncorrelated. On the IBM Quantum hardware Nation \textit{et al.} found this to be a reasonable assumption (using \textit{ibmq\_kolkata}), with little difference observed between this tensored approach versus a complete measurement calibration until inducing correlations by increasing the readout pulse amplitudes from their optimized values \cite{nation2021scalable}.

\begin{figure}[h]
    \centering
    \includegraphics[width=\linewidth]{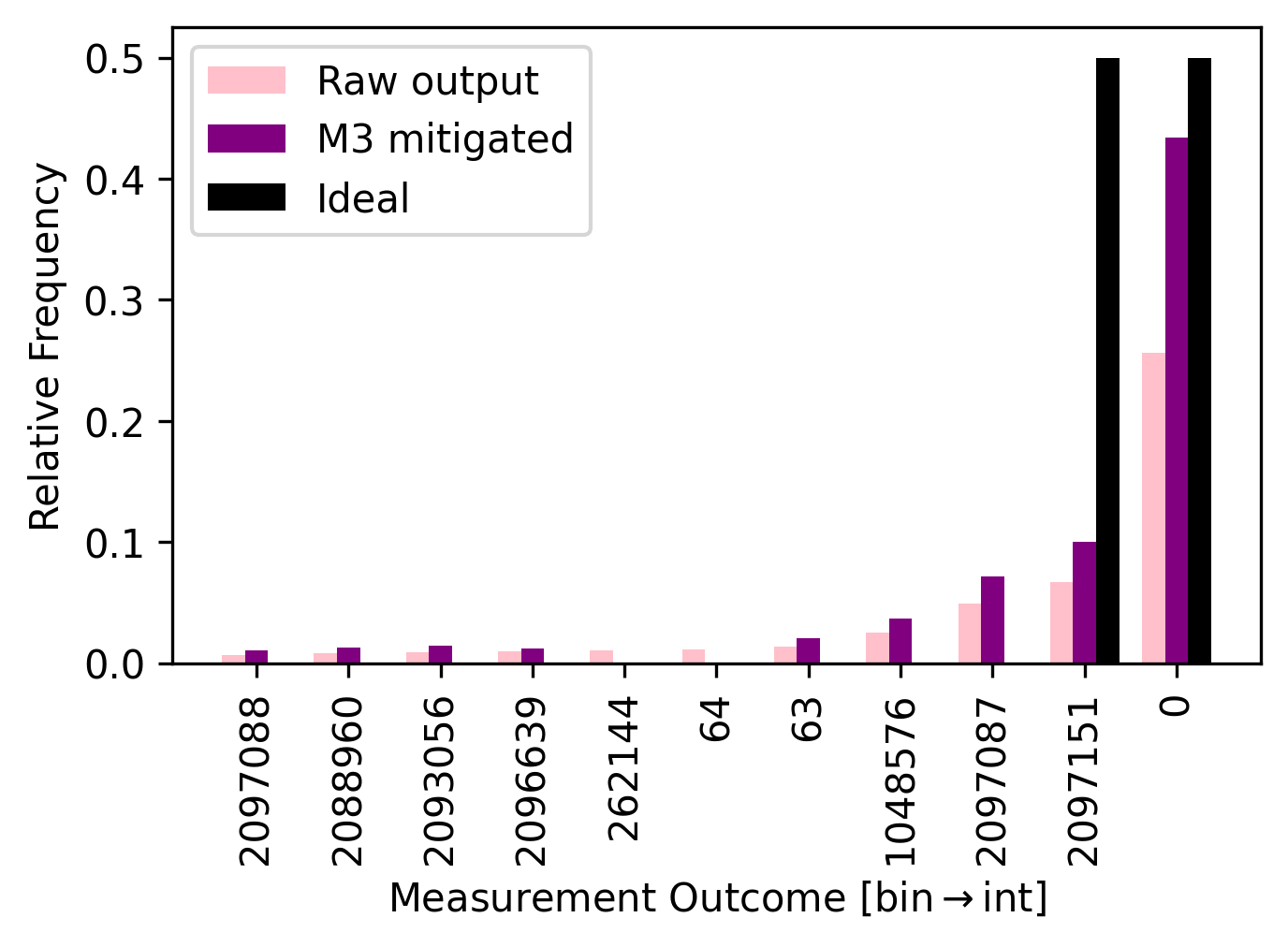}
    \caption{Comparing raw and MEM measurement distributions against the ideal output for 21-qubit GHZ preparation on \textit{ibmq\_kolkata} (the greatest number of qubits possible without SWAP operations) with $2^{15}$ circuit shots. Only outcomes exceeding a frequency of $10^{-2}$ are plotted here; this contributes $46.8\%$ and $71.4\%$ of the raw and MEM distributions, respectively. $T_1$ relaxation results in a reduced frequency of $\ket{\bm{1}}$ measurement outcomes compared with $\ket{\bm{0}}$ whereas they should be observed with equal probility $50\%$.}
    \label{fig:GHZ_21q_kolkata}
\end{figure}

The expression of $A$ in terms of single-qubit readout errors \eqref{sq_readout} requires just $2N$ quantum experiments to be carried out, versus $2^N$ in a complete measurement calibration. Furthermore, its form is particularly convenient as it is amenable to matrix-free iterative linear algebra techniques \cite{saad2003iterative}. The Python package \textit{mthree} developed through the work of Nation \textit{et al.} is available in Qiskit; we utilized this for our QEM benchmark and is the only technique presented here that we did not implement ourselves. In Figure \ref{fig:GHZ_21q_kolkata} we present the measurement distribution pre- and post-MEM for a 21-qubit GHZ preparation procedure on \textit{ibmq\_kolkata}, recalling from Figure \ref{fig:GHZ_benchmark} that we observed an increase from $29.3\%$ to $47.5\%$ in GHZ state fidelity. The effect of $T_1$ relaxation is also visible in this plot, whereby the $\ket{\bm{0}}$ state occurs with considerably greater probability than $\ket{\bm{1}}$ since the former is energetically favourable.

\subsection{Symmetry Verification}\label{sec:SV}

An inexpensive method of error mitigation is to take known symmetries of the Hamiltonian (usually those of the $\mathbb{Z}_2$ variety, i.e. Pauli operators that commute termwise across the Hamiltonian) and enforce stabilizer constraints on the measured binary strings resulting from a quantum experiment; we shall refer to this as \textit{symmetry verification} (SV) \cite{bonet2018low, mcardle2019error, cai2021quantum}. On the other hand, in Section \ref{sec:QR_techniques} we described how those same symmetries may instead be utilized for the purposes of qubit reduction, which allowed us to dramatically reduce the dimension of our Hamiltonian. In doing so, we may no longer use $\mathbb{Z}_2$ symmetries to postselect allowed measurement outcomes as the reduced Hamiltonian has been abstracted from them. However, there still exist symmetries of a more general nature that need not commute with each term individually, but do so with respect to the full Hamiltonian. Examples in the setting of electronic structure are the (Jordan-Wigner encoded) particle and spin quantum number operators
\begin{equation}\label{non_Z2_sym}
    S_{N} = \sum_{i=1}^N Z_i, \;\;
    S_{z} = \frac{1}{2} \sum_{i=1}^N (-1)^{i} Z_i;
\end{equation}
note how the latter differs from the up/down spin parity operators of \eqref{up_down_parity}. These are not $\mathbb{Z}_2$ symmetries as they do not commute with individual terms in the Hamiltonian and are therefore nontrivial in the contextual subspace; the projection procedure respects commutation and therefore we may use the reduced operators 
\begin{equation}
\begin{aligned}
    S_N ={} & 17 \cdot III - IIZ - \frac{1}{2} (IZI + IZZ + ZII + ZIZ), \\
    S_z ={} &  \frac{1}{4} ( IZI + IZZ - ZII - ZIZ )
\end{aligned} 
\end{equation}
for error mitigation in our \ce{HCl} 3-qubit contextual subspace -- as an exercise we suggest the reader confirms that these operators do indeed commute with the Hamiltonian described by the terms in Table \ref{tab:3q_hamiltonian}. An interesting feature of this reduced $S_N$ operator is the identity term that was not present in the original formulation of the number operator in \eqref{non_Z2_sym}; the coefficient indicates the number of particles that have been effectively projected out of the contextual subspace, in this case seventeen out of the eighteen available electrons. The rotations involved in the projection procedure abstract the reduced system from the underlying physical system, however this observation suggests there may be some natural interpretation of the contextual subspace method, which would be an interesting pursuit for further research.

An important point is that we may only mitigate errors of terms that commute with the number and spin operators which, in this case, means only the diagonal ones; this may still yield significant improvements in error since these terms have the greatest coefficient magnitude and errors here will be amplified proportionally.

Given an ensemble of measurements $\{\bm{b}\}$, we discard any binary strings $\bm{b} \in \mathbb{Z}_2^{N}$ that do not respect the number and spin symmetries; given that we know the number of particles $n$ in the system and the allowed spin values $\{s_0, \dots, s_{M-1}\}$ where $s_i = s-i$ for quantum number $s$ (multiplicity $M=2s+1$), we require that $S_N \ket{\bm{b}} = n \ket{\bm{b}}$ and $S_z \ket{\bm{b}} = s_i \ket{\bm{b}}$ for some $i \in \{0,\dots, M-1\}$. Our \ce{HCl} problem is in a singlet configuration, hence the only allowable spin value is $s=0$ and thus valid quantum measurements are those in the kernel of $S_z$.

This QEM technique requires no additional coherent overhead and only minor postprocessing, yet we observe respectable error suppression from enforcing number and spin symmetries on the diagonal Hamiltonian terms, as seen in Table \ref{mitigation_benchmark}. We intend to investigate the use of non-abelian point group symmetries (see Section \ref{sec:tapering}) for the purposes of error mitigation in future work, although it is not immediately clear whether this will be possible.  

\subsection{Zero-Noise Exptrapolation}\label{sec:ZNE}

The technique of \textit{zero-noise extrapolation} (ZNE), also referred to in the literature as \textit{richardson extrapolation}, operates on the principle that one may methodically amplify noise present in our quantum measurement output, obtaining a collection of increasingly noisy energy estimates before extrapolating the data and inferring the experimentally untouchable point of `zero noise' \cite{li2017efficient, temme2017error, endo2018practical, kandala2019error, giurgica2020digital, he2020zero, mari2021extending}. There are many methods of amplifying noise in our quantum circuits: some do so continuously by stretching gates temporally, requiring pulse-level control over the hardware, whereas others employ discrete approaches that either insert identity blocks of increasing complexity (e.g. unitary folding) or replace the target gate with a product over its roots. 

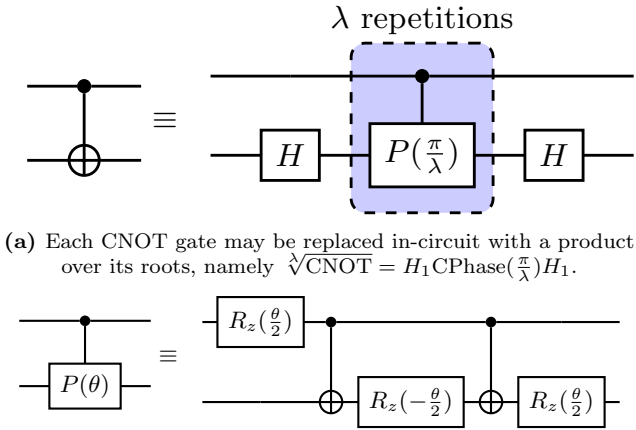
\begin{figure}[b]
    \centering
    \begin{subfigure}[b]{\linewidth}
         \centering
         \resizebox{\linewidth}{!}{
         \begin{quantikz}[row sep=0.5cm, column sep=0.4cm]
            \qw & \ctrl{1} & \qw \\
            \qw & \targ{} & \qw
        \end{quantikz} 
        $\equiv$
        \begin{quantikz}[row sep=0.4cm, column sep=0.5cm]
            \qw & \qw      & \ctrl{1}\gategroup[2,steps=1,style={dashed,rounded corners,fill=blue!20, inner xsep=2pt},background]{{\sc $\lambda$} repetitions}                  &  \qw      & \qw \\
            \qw & \gate{H} & \gate{P(\frac{\pi}{\lambda})} & \gate{H} & \qw
        \end{quantikz}
        }
         \caption{Each CNOT gate may be replaced in-circuit with a product over its roots, namely $\sqrt[\lambda]{\mathrm{CNOT}} = H_1 \mathrm{CPhase}(\frac{\pi}{\lambda}) H_1$.}
         \label{root_prod_circ}
     \end{subfigure}
     \hfill
     \begin{subfigure}[b]{\linewidth}
         \centering
         \begin{quantikz}[row sep=0.5cm, column sep=0.4cm]
            \qw & \ctrl{1} & \qw \\
            \qw & \gate{P(\theta)} & \qw
        \end{quantikz} 
        $\equiv$
         \begin{quantikz}[row sep=0.4cm, column sep=0.2cm]
            \qw & \gate{R_z(\frac{\theta}{2})} & \ctrl{1} & \qw                           & \ctrl{1} & \qw & \qw \\
            \qw & \qw                          & \targ{}  & \gate{R_z(-\frac{\theta}{2})} & \targ{}  & \gate{R_z(\frac{\theta}{2})} & \qw
        \end{quantikz}
         \caption{Since the IBM hardware takes the CNOT as its native entangling gate, the CPhase decomposition of \textbf{(a)} is transpiled back in terms of CNOTs at the point of execution.}
         \label{fig:CPhase_native}
     \end{subfigure}
    \caption{Noise amplification method used for zero-noise extrapolation. Given a noise amplification factor $\lambda \in \mathbb{N}$, each CNOT is replaced by $2\lambda$ CNOTs, $3\lambda$ single-qubit $Z$-rotations and two Hadamard gates.}
    \label{fig:noise_amp_method}
\end{figure}

It is the latter method we employ here. Given a quantum circuit $U$, some constituent native gate $G$ and a noise parameter $\lambda \in \mathbb{N}$, we shall replace each instance of $G$ in-circuit with the equivalent operation $\prod_{i=1}^\lambda \sqrt[\lambda]{G}$ to yield a noise-amplified circuit $U_\lambda$. One may note that $\lambda = 1$ corresponds with the unmodified circuit, whereas we intend to infer a value for $\lambda=0$ by evaluating expectation values $E_\lambda = \bra{\psi_{\mathrm{ref}}} U_\lambda^\dag P U_\lambda \ket{\psi_{\mathrm{ref}}}$ at integer values $\lambda \in \{1, 2, 3, \dots\}$ and extrapolating. 

In particular, we shall take $G = \mathrm{CNOT}$ since this is the dominant source of error by an order of magnitude, as seen in Table \ref{hardware_specs}. In order to decompose CNOT into its roots, we define the two-qubit gate
\begin{equation}
\begin{aligned}
    \mathrm{CPhase}(\theta) ={} & \frac{1}{2} \big[ (1+Z) \otimes I + (1-Z) \otimes P(\theta) \big] \\
    ={} & \begin{bmatrix}
        I & \bm{0} \\
        \bm{0} & P(\theta)
    \end{bmatrix}
\end{aligned}
\end{equation}
where $P(\theta) \coloneqq e^{i\theta/2} R_z(\theta) = \begin{bmatrix} 1 & 0 \\ 0 & e^{i\theta} \end{bmatrix}$ and note that $\mathrm{CNOT} = H_1 \cdot \mathrm{CPhase}(\pi) \cdot H_1$. In other words, the Hadamard gates applied on the target qubit diagonalize the $\mathrm{CNOT}$ gate and thus 
\begin{equation}
    \sqrt[\lambda]{\mathrm{CNOT}} = H_1 \cdot \mathrm{CPhase}\Big(\frac{\pi}{\lambda}\Big) \cdot H_1.
\end{equation}


\begin{figure}[b]
    \centering
    \includegraphics[width=\linewidth]{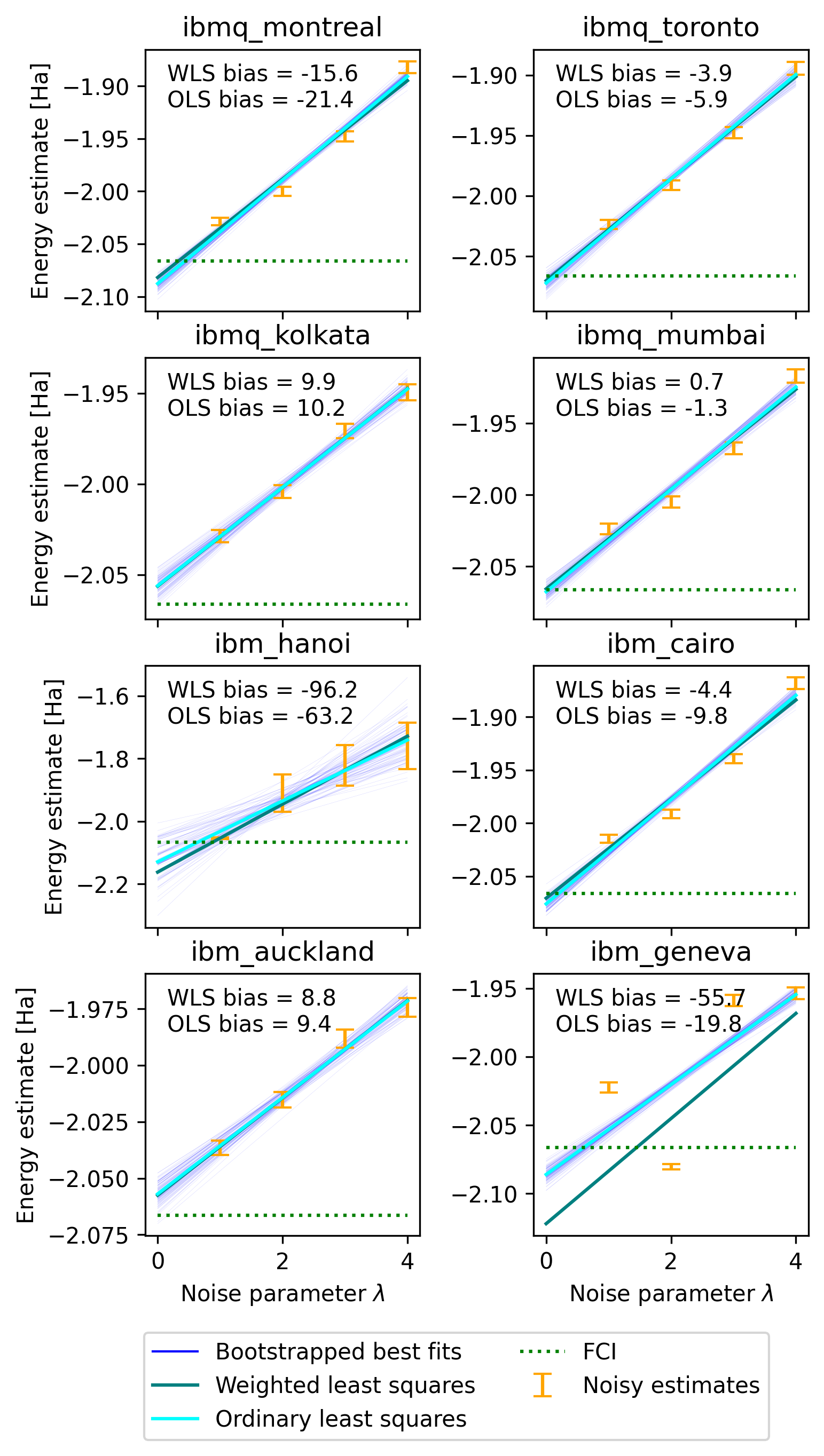}
    \caption{Zero-noise extrapolation of our \ce{HCl} problem, comparing weighted and ordinary least squares in addition to possible bootstrapped fits. Each of the noisy estimates have had measurement-error mitigation and symmetry verification applied.}
    \label{fig:ZNE_fits}
\end{figure}

The CNOT root-product decomposition is given as a circuit in Figure \ref{fig:noise_amp_method}\textbf{(a)}. When it comes down to implementation of zero-noise extrapolation on a quantum computer, one must be mindful of which gates are native to said device and should avoid circuit optimization routines since these may result in an unpredictable scaling of noise. For example, as stated in Section \ref{sec:hardware}, the CNOT is in fact the native entangling gate on IBM Quantum systems; therefore, CPhase operations will be transpiled back in terms of CNOT and $R_z$ gates at the point of execution, the decomposition of which is given in Figure \ref{fig:noise_amp_method}\textbf{(b)}. Such considerations can wreak havoc on zero-noise extrapolation if not controlled carefully.


For our specific implementation of ZNE we shall assume that the individual noise amplified estimates have been obtained via an estimator $\E{\lambda}$ so that $E_\lambda = \mathbb{E}(\E{\lambda})$, which might have previously had some other QEM strategy applied. We shall then evaluate estimates for $\lambda \in \{1,2,3,4\}$ before performing weighted least squares (WLS) regression with weights $w_\lambda = 1/\mathrm{var}(\mathcal{E}_{\lambda})$ to infer a `zero-noise' estimate $\E{ZNE} = E_0$. This penalises highly varying points in the extrapolation; in Figure \ref{fig:ZNE_fits} we compare WLS against ordinary least squares (OLS) and a bootstrapped collection of possible ZNE curves. We note that such a regression approach allows us to quantify the success of our extrapolation via the coefficient of determination, or $R^2$ value, expressed as a ratio of residual and total sum of squares \cite{draper1998applied}. WLS yields the smallest bias in all but two cases: \textit{ibm\_hanoi} and \textit{ibm\_geneva}. In the former we have a low-variance, low-bias point at $\lambda=1$ that is pinning the extrapolation whilst the noisier estimates vary dramatically, whereas the latter exhibits a problematic low-variance, negatively-biased point at $\lambda=2$ that is causing the extrapolation to fail.

\subsection{Dual-State Purification}\label{sec:DSP}

Purification-based error mitigation techniques operate on the basis that in quantum computation we are often interested in preparing some pure state $\ket{\psi_0}$, whereas in reality what is actually prepared on the noisy quantum hardware is some mixed state
\begin{equation}
    \rho = \sum_{i=0}^{2^N-1} \lambda_i \rho_i
\end{equation}
where $\rho_i = \ket{\psi_i} \bra{\psi_i}$ and we assume $\lambda_i > \lambda_j$ for $i<j$. The central observation that purification-based methods exploit is 
\begin{equation}\label{rho_M}
    \rho^M/\Tr{(\rho^M)} \rightarrow \rho_0 \;\; (M \rightarrow \infty),
\end{equation}
and the convergence is exponentially fast. This is precisely the formulation of \textit{virtual distillation} \cite{huggins2021virtual}, in which one prepares $M$ copies of the mixed state $\rho$ over disjoint quantum registers and induces their product via application of a cyclic shift operator. However, this permutation circuit is expensive and not feasible for near-term applications; the error mitigation technique we investigate here -- \textit{dual state purification} (DSP), also referred to in the literature as \textit{echo verification} \cite{cai2022quantum} -- is closely related but may be implemented at significantly reduced cost. While the technique was first presented in the context of quantum phase estimation (QPE) \cite{o2021error}, it was subsequently extended to the NISQ era \cite{cai2021resource, huo2022dual}. The idea behind this method is that one prepares some quantum state, performs an intermediary readout and subsequently uncomputes the circuit before postselecting on zero measurement outcomes; this bears some resemblance to second-order virtual distillation ($M=2$) but with the state
\begin{equation}
    (\rho \overline{\rho} + \overline{\rho} \rho)/2\Tr{(\rho \overline{\rho})}
\end{equation}
as opposed to form given in \eqref{rho_M} \cite{huo2022dual}. 

We now describe explicitly the steps one must follow to implement DSP. The setting is that of a Pauli operator $P \in \mathcal{P}_N$ whose expectation value we wish to evaluate with respect to an $N$-qubit state $\ket{\psi} = U \ket{\bm{0}}$. Denoting by $\mathcal{I}$ the set of non-identity qubit indices we may identify a change-of-basis operator $B$ such that $B P B^\dag = Z_\mathcal{I} = \otimes_{i \in \mathcal{I}} Z_i$ defined as
\begin{equation}
    B_i = \begin{cases}
        I,  & P_i \in \{I, Z\} \\
        H,  & P_i = X \\
        HS, & P_i = Y.
    \end{cases}
\end{equation}
Now, we note the effect of applying a $\mathrm{CNOT}$ gate controlled on a qubit position $i \in \mathcal{I}$ to an ancilla register. With the expression
\begin{equation}
    \mathrm{CNOT} = \frac{1}{2} \big[ (I+Z) \otimes I + (I-Z) \otimes X \big]
\end{equation}
we observe
\begin{equation}
\begin{aligned}
    \mathrm{CNOT}_{i,a} \big(\ket{\psi} \otimes \ket{0}_{a}\big) ={} & 
    \frac{1}{2} \big( \ket{\psi} \otimes \ket{0}_{a} + Z_i \ket{\psi} \otimes \ket{0}_{a} + \\ 
    {} & \hspace{4.5mm} \ket{\psi} \otimes \ket{1}_{a} - Z_i \ket{\psi} \otimes \ket{1}_{a} \big) \\
    ={} & \frac{1}{\sqrt{2}} \big( \ket{\psi} \otimes \ket{+}_a + Z_i \ket{\psi} \otimes \ket{-}_a \big).
\end{aligned}
\end{equation}
Finally, as demonstrated by Huo \& Li \cite{huo2022dual}, we may uncompute the circuit $U$ that prepares $\ket{\psi}$ and post-select on measurement outcomes $\ket{\bm{0}}$, occurring with probability $p_{\bm{0}}$, to drive the ancilla register into the state $\frac{1}{\sqrt{2 p_{\bm{0}}}} \big( \ket{+}_a + \braket{\psi | Z_i | \psi} \ket{-}_a \big)$.

We now describe the full process of computing the expectation value $\braket{P}_{\psi}$. First of all, the circuit is initialized in the state
\begin{equation}
    \ket{\psi_0} = \ket{\bm{0}} \otimes \ket{0}_a
\end{equation}
before applying the unitary $U$ and basis transformation $B$ supported on some qubit subset $\mathcal{I}$:
\begin{equation}
\begin{aligned}    
    \ket{\psi_1} 
    ={} & \big(BU \otimes I \big) \ket{\psi_0} \\
    ={} & BU \ket{\bm{0}} \otimes \ket{0}.
\end{aligned}
\end{equation}
We now compute and store the parity of qubits $\mathcal{I}$ on the ancilla register:
\begin{equation}
\begin{aligned}    
    \ket{\psi_2} 
    ={} & \prod_{i \in \mathcal{I}} \mathrm{CNOT}_{i,a} \ket{\psi_1} \\
    ={} & \frac{1}{\sqrt{2}} \Big( BU\ket{\bm{0}} \otimes \ket{+}_a + Z_{\mathcal{I}} B U \ket{\bm{0}} \otimes \ket{-}_a  \Big).
\end{aligned}
\end{equation}
Inverting the change-of-basis and unitary circuit we obtain
\begin{equation}
\begin{aligned}    
    \ket{\psi_3} 
    ={} & \big(U^\dag B^\dag \otimes I \big) \ket{\psi_2} \\
    ={} & \frac{1}{\sqrt{2}} \Big( \ket{\bm{0}} \otimes \ket{+}_a + U^\dag \underbrace{B^\dag Z_{\mathcal{I}} B}_{=P} U \ket{\bm{0}} \otimes \ket{-}_a  \Big) \\
    ={} & \frac{1}{\sqrt{2}} \Big( \ket{\bm{0}} \otimes \ket{+}_a + U^\dag P U \ket{\bm{0}} \otimes \ket{-}_a  \Big).
\end{aligned}
\end{equation}
Finally, we perform a projective measurement onto the $\bm{0}$ outcome, effected by the projection operator $P_{\bm{0}} = \ket{\bm{0}}\bra{\bm{0}}$ with probability $p_{\bm{0}} = \bra{\psi} P_{\bm{0}} \ket{\psi}$:
\begin{equation}\label{ancilla_state}
\begin{aligned}    
    \ket{\psi_4} 
    ={} & \frac{1}{\sqrt{p_{\bm{0}}}} \big(P_{\bm{0}} \otimes I \big) \ket{\psi_3} \\
    ={} & \frac{1}{\sqrt{2 p_{\bm{0}}}} \Big( \ket{\bm{0}} \otimes \ket{+}_a + \ket{\bm{0}}\underbrace{\bra{\bm{0}} U^\dag P U \ket{\bm{0}}}_{= \braket{P}_{\psi}} \otimes \ket{-}_a  \Big) \\
    ={} & \ket{\bm{0}} \otimes \underbrace{\frac{1}{\sqrt{2 p_{\bm{0}}}} \Big( \ket{+}_a  + \braket{P}_{\psi} \ket{-}_a \Big)}_{:=\ket{\phi}_a};
\end{aligned}
\end{equation}
in practice, this projective measurement is realised by post-selecting on zero measurement outcomes.

In effect, we have induced a virtual calculation of the desired expectation value on the ancilla qubit. The quantity $\braket{P}_{\psi}$ may be extracted by performing measurements of the ancilla state $\ket{\phi}$ in the $X$ and $Z$ bases, as we shall demonstrate now.

Using the normalization condition for $\ket{\phi}$ we infer that
\begin{equation}
    p_{\bm{0}} = \frac{1 + \braket{P}_{\psi}^2}{2},
\end{equation}
which we note is at least $\frac{1}{2}$, meaning we should in principle retain at worst $50\%$ of the samples taken from the quantum hardware. Inspecting \eqref{ancilla_state}, sampling from the $\ket{\phi}$ state in the $X$-basis yields $\ket{+}$ and $\ket{-}$ with probabilities $p^X_0$ and $p^X_1$, respectively, from which we obtain the estimator
\begin{equation}\label{X_basis}
    \E{}^X
    = p^X_0- p^X_1 
    \approx \frac{1-\braket{P}_{\psi}^2}{2 p_{\bm{0}}}
    = \frac{1-\braket{P}_{\psi}^2}{1+\braket{P}_{\psi}^2}.
\end{equation}
for the ancilla expectation value $\braket{X}_{\phi} $.

We may also express $\ket{\phi}$ in the $Z$-basis 
\begin{equation}
    \ket{\phi} = \frac{1}{2 \sqrt{p_{\bm{0}}}} \Big[ \big(1+\braket{P}_{\psi}\big) \ket{0}_a + \big(1-\braket{P}_{\psi}\big) \ket{1}_a \Big];
\end{equation}
sampling from this state we obtain $\ket{0}$ and $\ket{1}$ with probabilities $p^Z_0$ and $p^Z_1$, respectively. From this we may derive an estimator
\begin{equation}\label{Z_basis}
\begin{aligned}
        \E{}^Z
        ={} & p^Z_0 - p^Z_1 \\
        \approx{} & \frac{1}{4 p_{\bm{0}}} \Big[ \big(1+\braket{P}_{\psi}\big)^2 - \big(1-\braket{P}_{\psi}\big)^2 \Big] \\
        ={} & \frac{\braket{P}_{\psi}}{p_{\bm{0}}} \\
        ={} & \frac{2 \braket{P}_{\psi}}{1+\braket{P}_{\psi}^2}
\end{aligned}
\end{equation}
for the ancilla expectation value $\braket{Z}_{\phi} $.

Finally, by combining \eqref{X_basis} and \eqref{Z_basis} we may reconstruct an error mitigated estimator for the desired quantity $\braket{P}_{\psi}$:
\begin{equation}\label{expval_P}
     \E{DSP} = \frac{\E{}^Z}{1+\E{}^X}.
\end{equation}
One may actually reconstruct $\braket{P}_{\psi}$ using \textit{only} the $Z$-basis measurements by noting $\braket{X}_{\phi}^2 + \braket{Z}_{\phi}^2 \equiv 1$ and therefore
\begin{equation}
    \braket{P}_{\psi} = \frac{\braket{Z}_{\phi}}{1+\sqrt{1-\braket{Z}_{\phi}}},
\end{equation}
which one may arrive at by forming a quadratic equation from \eqref{Z_basis} and solving. Doing the same for the $X$-basis measurements yields
\begin{equation}
    \braket{P}_{\psi} = \pm \sqrt{\frac{1-\braket{X}_{\phi}}{1+\braket{X}_{\phi}}},
\end{equation}
however it is not possible to determine the correct sign using these measurements alone; supplementary $Z$-basis measurements would be required to indicate the sign here.

One might also note that, expressing $\ket{\phi}$ in the $Y$-basis
\begin{equation}
\begin{aligned}
    \ket{\phi} = \frac{1}{2 \sqrt{2 p_{\bm{0}}}}
    \bigg\{ & \Big[ \big(1+\braket{P}_{\psi}\big) - i \big(1-\braket{P}_{\psi}\big) \Big] \ket{+_i}_a \\ 
    + & \Big[ \big(1+\braket{P}_{\psi}\big) + i \big(1-\braket{P}_{\psi}\big) \Big] \ket{-_i}_a \bigg\},
\end{aligned}
\end{equation}
we must have $\braket{Y}_{\phi} = p^Y_0 - p^Y_1 = 0$; this was also noted by Huo \& Li \cite{huo2022dual} and we might be able to exploit this fact for additional error mitigation in future work.

\begin{figure}[h]
    \centering
    \input{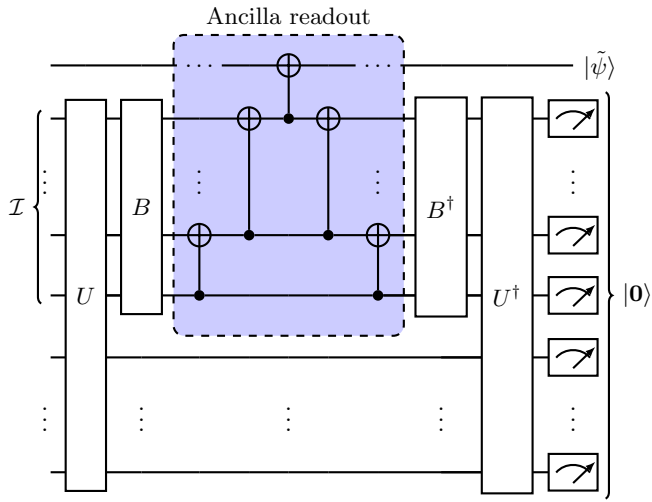}
    \caption{Schematic of the dual-state purification protocol with readout to a single ancilla qubit, where $\mathcal{I}$ indicates the non-identity qubit positions of the Hamiltonian term being measured. In reality, one must construct the readout sub-circuit with careful consideration of the chip topology to avoid excessive SWAP usage; for the 3-qubit \ce{HCl} problem we report the optimal readout blocks in Figure \ref{fig:readout_map}.}
    \label{fig:dsp_circuit}
\end{figure}

In Figure \ref{fig:dsp_circuit} we present the DSP circuit. The only errors that are not suppressed through this process are those occurring in the readout phase, since errors may propagate through to the ancilla register and are not cancelled during the subsequent uncomputation. However, there is one additional trick we may employ here; if the circuit is error-free, then the state of the ancilla qubit is necessarily pure. In practice, the ancilla will be described by a mixed state 
\begin{equation}
    \rho = (1-\epsilon) \ket{\varphi_0}\bra{\varphi_0} + \epsilon \ket{\varphi_1}\bra{\varphi_1}    
\end{equation}
where $\epsilon$ is the infidelity,
which we may characterise fully via state tomography. Measuring the ancilla in the $X, Y, Z$ bases we may reconstruct $\rho = \frac{1}{2} ( I + \gamma_X X + \gamma_Y Y + \gamma_Z Z )$ where $\gamma_P = \Tr{(P \rho)}$ and identifying the largest eigenvalue with corresponding eigenvector $\ket{\varphi_0}$ we take this as an approximation to the pure state $\ket{\phi}$ obtained in the noiseless setting. Huo \& Li \cite{huo2022dual} found this additional state tomography procedure to be essential in obtaining accurate results from dual-state purification.

Furthermore, from \eqref{X_basis} we note that $\braket{X}_{\phi} \geq 0$, but in practice it is possible for negative value to appear from quantum experiments. In fact, the depolarizing noise can be sufficiently high such that the corresponding eigenvalue of $\rho$ dominates, resulting in spurious expectation values that can violate this non-negativity constraint maximally. This is a considerable problem when one considers the form \eqref{expval_P}, since this can result in division by zero, yielding a potentially infinite expectation value estimate for $\braket{P}_{\psi}$. We combat this by always choosing the eigenvalue with positive $\braket{X}_{\phi}$, even in the case when it does not hold the greatest weight. We observed this in particular for terms necessitating expensive SWAP operations; for our \ce{HCl} circuit this meant only terms of the form $ZZI$ under change-of-basis (see Section \ref{sec:ancilla_readout} for details), since this results in a closed loop of three CNOT gates which is not directly expressible on any IBM system (the heavy-hex topology of Figure \ref{fig:kolkata_topology} does not contain cycles of three connected qubits). We also observed instability of the tomography purification method when $\braket{Z}_{\phi} \approx 0$ whereby the error can be increased through this procedure. Therefore, we opted only to run this additional step when the raw expectation value exceeded some threshold near zero, taking the standard DSP result otherwise.

A potential modification for future work would be to flip the initial state of the system register $\ket{\bm{0}} \rightarrow \ket{\bm{1}}$ via a layer of $X$ gates and postselect on $\bm{1}$ measurement outcomes. While this should theoretically be no different to initializing with $\ket{\bm{0}}$, the effect of $T_1$ relaxation is for qubits to decay into the energetically favourable $\ket{0}$ state (as was observed in Figure \ref{fig:GHZ_21q_kolkata}), resulting in the erroneous postselection of invalid measurements. By flipping the initial state, we should expect to retain fewer measurements in the postselected data, but the probability of these corresponding with successful circuit runs should be improved.


\section{Ground State Preparation}\label{GS_prep}

Before proceeding onto the quantum error mitigation benchmark, there are a few additional considerations to resolve. Firstly, one must identify a suitable ansatz circuit that is sufficiently expressible to realize the desired ground state. Secondly, we discuss the mapping of our circuits onto physical qubits, in particular for dual-state purification since one should be mindful of the added qubit connectivity constraints arising from parity computation stored on the ancilla qubit. Thirdly, despite not implementing any shot reduction methods in this work, we still wish to distribute the shot budget in an informed manner, preferably tailored to each device; this is the final point of discussion before moving onto the results of our benchmark.

\subsection{Ansatz Construction}

Initially, we tested the noncontextual projection ansatz \cite{weaving2023stabilizer} derived from the 316-term CCSD operator. The projection into the 3-qubit contextual subspace yields a 6-term excitation pool from which we identify 4 operators via qubit-ADAPT-VQE that permit chemical precision. Despite this dramatic reduction in circuit depth from the full UCCSD ansatz, the resulting noncontextual projection ansatz consists of 12 CNOT gates which we found to be prohibitive in achieving chemical precision.

\begin{figure}[h]
    \centering
    \resizebox{0.9\linewidth}{!}{
    \input{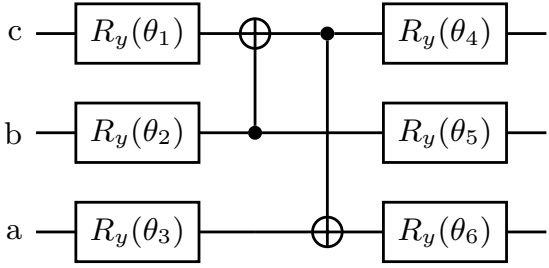}
    }
    \caption{Hardware efficient \ce{HCl} $3$-qubit contextual subspace ansatz; the $Y$-rotation gates are decomposed into native gates as $R_y = \sqrt{X} R_z \sqrt{X}$. The optimal parametrization obtained from the statevector simulation in Figure \ref{fig:noiseless_sim} is: $\theta_1 = -0.06492667,  \theta_2 = 2.89836152,  \theta_3 = 0.26373807,$ $\theta_4 = -0.06709062,  \theta_5 = 0.01006833, \theta_6 = -0.26585046$.}
    \label{fig:3q_ansatz}
\end{figure}

\begin{figure}[h!]
    \centering
    \includegraphics[width=\linewidth]{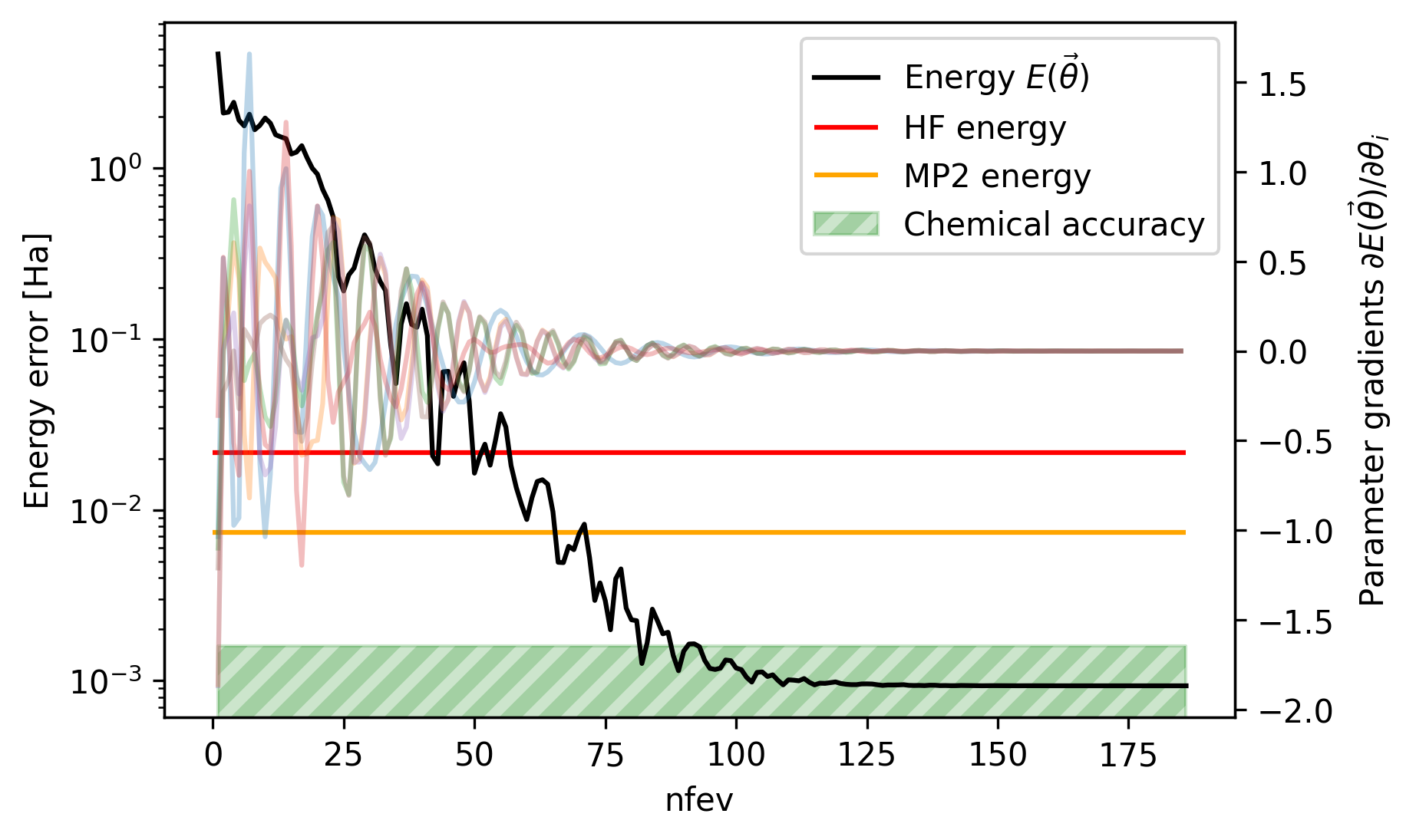}
    \caption{Noiseless 3-qubit CS-VQE simulation of the \ce{HCl} molecule over the hardware efficient ansatz presented in Figure \ref{fig:3q_ansatz}. The classical optimizer used is Adaptive Moment (Adam) estimation with gradients calculated using the parameter shift rule; we see that the ansatz is sufficiently expressible to achieve chemical accuracy.}
    \label{fig:noiseless_sim}
\end{figure}

To remedy this, we abandon chemical intuition in the name of hardware efficiency. It is already known that an arbitrary 3-qubit quantum state may be prepared on quantum hardware using at most 4 CNOT gates \cite{vznidarivc2008optimal}. In fact, we found that only 2 CNOT gates are sufficient in constructing a 3-qubit ansatz circuit that is sufficiently expressible for our electronic structure problem, presented in Figure \ref{fig:3q_ansatz}. In Figure \ref{fig:noiseless_sim} we present the outcome of a noiseless VQE simulation over this ansatz to illustrate its expressibility. 

\subsection{Ancilla Readout Mapping for DSP}\label{sec:ancilla_readout}

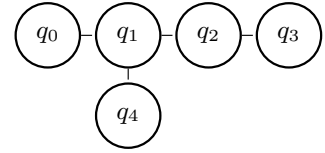
\begin{figure}[h]
    \centering
    \resizebox{0.5\linewidth}{!}{
    \begin{tikzpicture}[shorten >=1pt, auto, node distance=10mm,
     every node/.style={circle,thick,draw,minimum size=8mm}
     ]
    \node (A)              {$q_0$};    \node (B) [right of=A] {$q_1$};    \node (C) [right of=B] {$q_2$}; \node (D) [right of=C] {$q_3$};
    \node (E) [below of=B] {$q_4$}; 
    \draw (A) edge (B); \draw (B) edge (C); \draw (C) edge (D); \draw (B) edge (E);
    \end{tikzpicture}
    }
    \caption{The five-qubit cluster we require for dual-state purification in order to facilitate readout in every possible measurement basis. There are 18 such clusters on the 27-qubit Falcon chip (see Figure \ref{fig:kolkata_topology}) and we selected the optimal one with respect to gate and readout errors.}
    \label{fig:dsp_cluster}
\end{figure}

\begin{figure}[h!]
\centering

    \begin{subfigure}[t]{.3\linewidth}
    \begin{quantikz}[row sep=0.2cm]
        \lstick{$d$} & \targ{}   & \qw  \\
        \lstick{$c$} & \ctrl{-1} & \qw  \\
        \lstick{$b$} & \qw       & \qw  \\
        \lstick{$a$} & \qw       & \qw
    \end{quantikz}
    \caption{$IIZ$ \hspace{\textwidth} $a,b,c,d \mapsto q_0,q_2,q_1,q_4$}\label{fig:IIZ_readout}
    \end{subfigure}
    \begin{subfigure}[t]{.3\linewidth}
    \begin{quantikz}[row sep=0.2cm]
        \lstick{$d$} & \targ{}   & \qw  \\
        \lstick{$c$} & \qw       & \qw  \\
        \lstick{$b$} & \ctrl{-2} & \qw  \\
        \lstick{$a$} & \qw       & \qw
    \end{quantikz}
    \caption{$IZI$ \hspace{\textwidth} $a,b,c,d \mapsto q_0,q_2,q_1,q_3$}\label{fig:IZI_readout}
    \end{subfigure}
    \begin{subfigure}[t]{.3\linewidth}
    \begin{quantikz}[row sep=0.2cm]
        \lstick{$d$} & \targ{}   & \qw  \\
        \lstick{$c$} & \qw       & \qw  \\
        \lstick{$b$} & \qw       & \qw  \\
        \lstick{$a$} & \ctrl{-3} & \qw
    \end{quantikz}
    \caption{$ZII$ \hspace{\textwidth} $a,b,c,d \mapsto q_1,q_3,q_2,q_4$}\label{fig:ZII_readout}
    \end{subfigure}
    
    \begin{subfigure}[t]{.45\linewidth}
    \begin{quantikz}[row sep=0.2cm]
        \lstick{$d$} & \qw       & \targ{}   & \qw  \\
        \lstick{$c$} & \targ{}   & \ctrl{-1} & \qw  \\
        \lstick{$b$} & \ctrl{-1} & \qw       & \qw  \\
        \lstick{$a$} & \qw       & \qw       & \qw
    \end{quantikz}
    \caption{$IZZ$ \hspace{\textwidth} $a,b,c,d \mapsto q_0,q_2,q_1,q_4$}\label{fig:IZZ_readout}
    \end{subfigure}
    \begin{subfigure}[t]{.45\linewidth}
    \begin{quantikz}[row sep=0.2cm]
        \lstick{$d$} & \qw       & \targ{}   & \qw  \\
        \lstick{$c$} & \targ{}   & \ctrl{-1} & \qw  \\
        \lstick{$b$} & \qw       & \qw       & \qw  \\
        \lstick{$a$} & \ctrl{-2} & \qw       & \qw
    \end{quantikz}
    \caption{$ZIZ$ \hspace{\textwidth} $a,b,c,d \mapsto q_0,q_2,q_1,q_4$}\label{fig:ZIZ_readout}
    \end{subfigure}

    \begin{subfigure}[t]{.45\linewidth}
    \begin{quantikz}[row sep=0.2cm]
        \lstick{$d$} & \qw       & \qw       & \targ{}   & \qw  \\
        \lstick{$c$} & \targX{}   & \targ{}   & \ctrl{-1} & \qw  \\
        \lstick{$b$} & \qw       & \ctrl{-1} & \qw       & \qw  \\
        \lstick{$a$} & \swap{-2} & \qw       & \qw       & \qw
    \end{quantikz}
    \caption{$ZZI$ \hspace{\textwidth} $a,b,c,d \mapsto q_0,q_2,q_1,q_4$}\label{fig:ZZI_readout}
    \end{subfigure}
    \begin{subfigure}[t]{.45\linewidth}
    \begin{quantikz}[row sep=0.2cm]
        \lstick{$d$} & \qw       & \qw       & \targ{}   & \qw  \\
        \lstick{$c$} & \targ{}   & \targ{}   & \ctrl{-1} & \qw  \\
        \lstick{$b$} & \qw       & \ctrl{-1} & \qw       & \qw  \\
        \lstick{$a$} & \ctrl{-2} & \qw       & \qw       & \qw
    \end{quantikz}
    \caption{$ZZZ$ \hspace{\textwidth} $a,b,c,d \mapsto q_0,q_2,q_1,q_4$}\label{fig:ZZZ_readout}
    \end{subfigure}

\caption{Ancilla readout mappings given qubit clusters of the form in Figure \ref{fig:dsp_cluster}. Given the qubit topology of Figure \ref{fig:kolkata_topology} and the form of our ansatz in Figure \ref{fig:3q_ansatz} (where qubits $a,b,c$ are the same as above with $d$ the ancilla qubit), we may not entangle qubits $a$ and $b$ since it would result in a closed loop of three CNOT operations that is not expressible on the available quantum systems. We avoid this situation by introducing a single SWAP operation (represented in-circuit as \includegraphics[width=7mm]{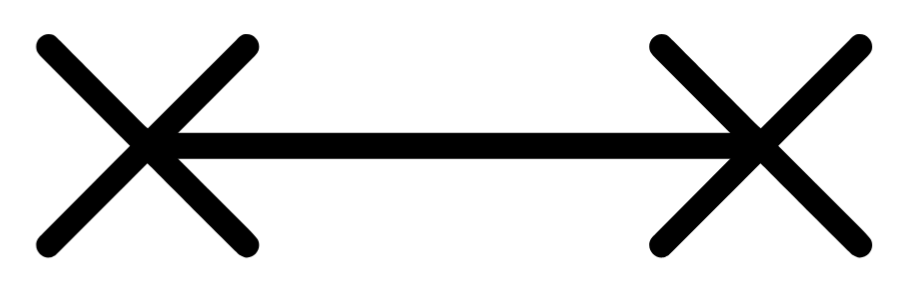}) for Hamiltonian terms of the form $ZZI$ as in \textbf{(f)}.}
\label{fig:readout_map}
\end{figure}

The main bottleneck for dual-state purification is the ancilla readout step. Given the limited topology of the available quantum systems (Figure \ref{fig:kolkata_topology}) and the structure of our Ansatz (Figure \ref{fig:3q_ansatz}), it is not possible to realize every 3-qubit Pauli $Z$ measurement basis ($IIZ, IZI, IZZ, ZII, ZIZ, ZZI, ZZZ$) without the aid of SWAP operations since at least one basis will always result in a closed loop of three CNOTs, which cannot be directly implemented on the hardware. We identified an optimal readout mapping that ensures just one measurement basis requires a SWAP operation by selecting a cluster of five qubit of the form in Figure \ref{fig:dsp_cluster} and implementing the readout as per Figure \ref{fig:readout_map}.

\subsection{Shot Budget Distribution}

To ensure a fair comparison, we define a fixed shot budget $B$ up front and distribute according to the particular combined error-mitigation strategy. The optimal shot distribution is in proportion with $v_P \coloneqq |h_P| \sqrt{\mathrm{var}(P)}$ where $\mathrm{var}(P) = 1 - \braket{P}_\psi^2$ \cite{rubin2018application}; however, the state-dependency means this may only be evaluated in-circuit. Therefore, we allocate $0.1\% (b = 0.001)$  of the overall budget to determine a rough estimate of the variance for each Hamiltonian term in order to rebalance the shot distribution accordingly; after this preliminary step we are left with $B^\prime = (1-b)B$ remaining shots. For example, defining $V = \sum_P v_P$ we allow
\begin{enumerate}
    \item ZNE: $\frac{B^\prime v_P}{\Lambda V}$ circuit shots for each Pauli term $P$ per noise amplification factor where $\Lambda$ is the number of noisy estimates desired for the energy extrapolation procedure.
    \item DSP: $\frac{B^\prime v_P}{2 V}$ circuit shots for each Pauli term $P$, where the factor of $\frac{1}{2}$ comes from performing both $X$ and $Z$ measurements over the ancilla qubit.
    \item DSP+ZNE: $\frac{B^\prime v_P}{2 \Lambda V}$ circuit shots for each Pauli term $P$ per noise amplification factor.
\end{enumerate}
Since the shot budget is fixed, layering multiple error-mitigation techniques may result in increased variance since fewer shots might be allocated to individual point estimates. It is the goal of this work to practically evaluate this trade-off between absolute error and uncertainty in the energy estimate, which has been noted in numerous studies \cite{takagi2022fundamental, cai2022quantum}.

\subsection{Methods}
To construct the  molecular Hamiltonian for \ce{HCl} (bond length 1.341 \AA), we  first performed a restricted Hartree-Fock calculation in \textit{PySCF} \cite{sun2018pyscf} in the STO-3G basis. \textit{OpenFermion} was then used to build the second quantised fermionic molecular Hamiltonian \cite{mcclean2020openfermion} and was mapped onto Pauli operators via the Jordan-Wigner transformation \cite{jordan1993paulische}. This was then converted into the \textit{Symmer} \cite{symmer2022} operator representation to leverage the included tapering and contextual subspace functionality, which facilitated a reduction to 3-qubits while incurring a ground state energy error of just $0.837$ mHa in the resulting contextual subspace Hamiltonian with respect to full configuration interaction (FCI); Section \ref{sec:QR_techniques} discusses this in further detail. 

We used \textit{Qiskit} \cite{Qiskit} for the construction of our hardware efficient ansatz circuit and the state preparation jobs required for each quantum error mitigation (QEM) strategy were composed as \textit{Qiskit Runtime} programs. These were submitted to the \textit{IBM Quantum} service and allowed us to retrieve all the necessary quantum circuit samples in the shortest amount of time possible to mitigate against noise drift.

The \textit{mthree} \cite{nation2021scalable} package was utilized to perform measurement-error mitigation (MEM, Section \ref{sec:MEM}) whereas we wrote bespoke implementations for all the other QEM techniques introduced in Section \ref{sec:QEM}, namely symmetry verifcation (SV, Section \ref{sec:SV}), zero-noise extrapolation (ZNE, Section \ref{sec:ZNE}) and dual-state purification (DSP, Section \ref{sec:DSP}) with or without tomography purification (TP). The linear regression functionality of \textit{statsmodels} \cite{seabold2010statsmodels} was utilized for the purposes of ZNE and the relevant post-processing required for each QEM technique was parallelized with \textit{multiprocessing} to permit a greater number of resamples to be extracted in our bootstrapping procedures (see Appendix \ref{sec:bootstrapping} for details).

We also provide all the Hamiltonian data, runtime program scripts, quantum experiment data and post-processing functions to aid the reader in reproducing the results of this paper, accessible via GitHub \cite{Weaving2023QEM}.

\subsection{Results}\label{sec:results}

\begin{table}[b]
\centering
\caption{Average error suppression and change in standard deviation under each error mitigation strategy evaluated across all 27-qubit Falcon IBM Quantum devices (excluding ibm\_hanoi and ibm\_geneva, which did not perform well as seen in the ZNE plots of Figure \ref{fig:ZNE_fits} and Table \ref{mitigation_benchmark}). Ordered by decreasing mean error suppression.}
\label{err_suppression}
\begin{tabular}{lrrrrrr}
\toprule
{} & \multicolumn{3}{l}{Error Suppression [\%]} & \multicolumn{3}{l}{Change in Std Dev} \\
{} &                  Mean &   Best &    Worst &              Mean &  Best &  Worst \\
\midrule
MEM+SV\\+ZNE    &                94.327 & 99.392 &   88.101 &             3.680 & 1.078 &  7.207 \\
DSP+TP         &                93.253 & 99.713 &   80.793 &             2.543 & 1.583 &  3.833 \\
MEM+DSP\\+TP     &                92.661 & 98.601 &   75.508 &             2.113 & 0.789 &  3.472 \\
MEM+ZNE        &                87.094 & 97.877 &   69.185 &             7.069 & 1.202 & 25.063 \\
MEM+SV        &                82.678 & 96.643 &   67.108 &             0.638 & 0.519 &  0.758 \\
SV+ZNE        &                79.799 & 94.938 &   52.882 &             4.270 & 0.385 &  8.594 \\
MEM            &                76.505 & 96.704 &   65.358 &             0.669 & 0.517 &  0.762 \\
SV            &                63.577 & 80.992 &   33.191 &             0.748 & 0.645 &  0.887 \\
MEM+DSP\\+TP+ZNE &                59.767 & 99.758 &  -98.987 &             6.738 & 3.104 &  8.853 \\
DSP+TP\\+ZNE     &                34.012 & 95.721 & -107.805 &             7.462 & 5.593 &  9.523 \\
ZNE            &                33.384 & 52.230 &   18.303 &             6.699 & 0.642 & 24.689 \\
MEM+DSP\\+ZNE    &               -10.180 & 93.343 & -238.874 &             6.779 & 3.715 &  8.601 \\
MEM+DSP        &               -18.002 & 75.430 & -330.030 &             2.224 & 1.080 &  3.440 \\
DSP+ZNE        &               -68.019 &  1.854 & -298.966 &             7.235 & 5.453 &  8.989 \\
DSP            &               -76.967 & 29.366 & -393.687 &             2.620 & 1.896 &  3.726 \\
\bottomrule
\end{tabular}
\end{table}

\begin{figure*}
\centering

    \begin{subfigure}[t]{.45\linewidth}
    \includegraphics[width=\linewidth]{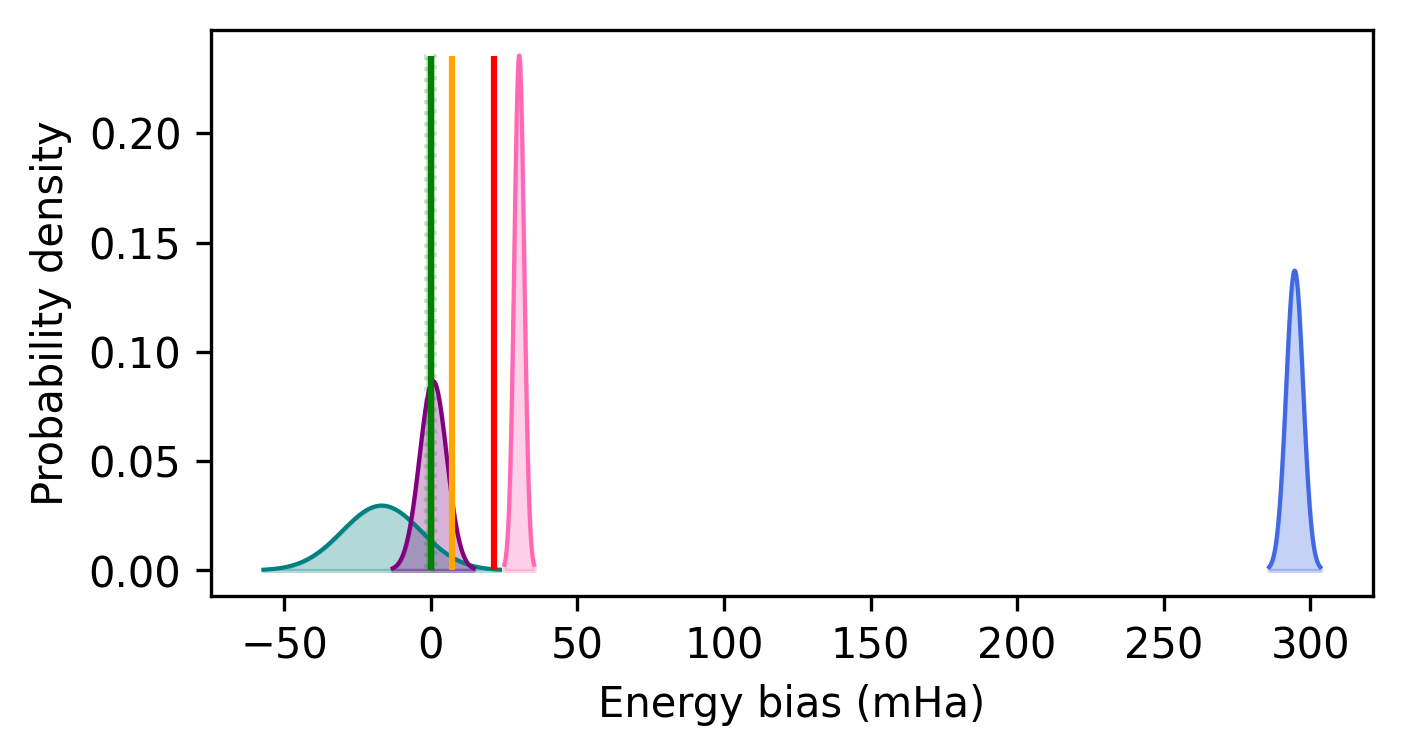}
    \caption{ibmq\_montreal}\label{fig:montreal_hist}
    \end{subfigure}
    \begin{subfigure}[t]{.45\linewidth}
    \includegraphics[width=\linewidth]{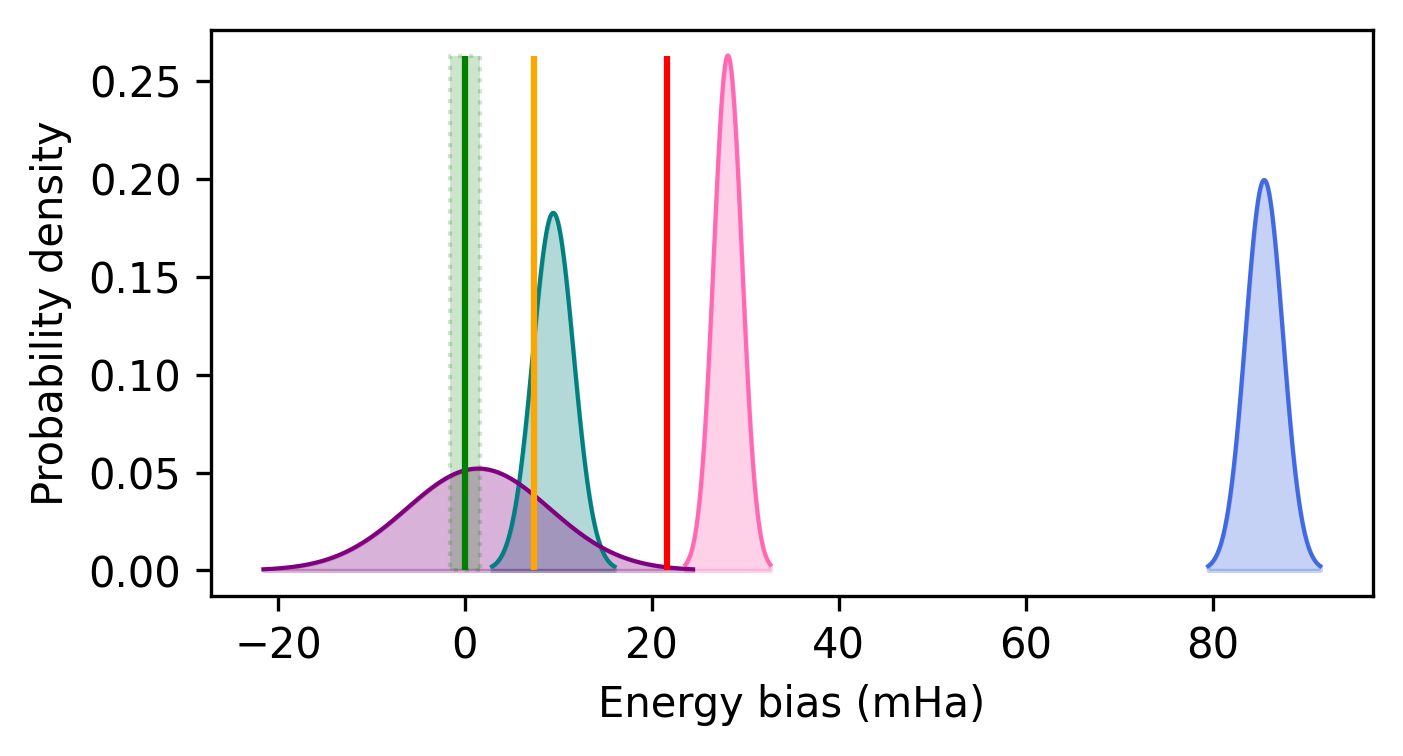}
    \caption{ibmq\_kolkata}\label{fig:kolkata_hist}
    \end{subfigure}
    \begin{subfigure}[t]{.45\linewidth}
    \includegraphics[width=\linewidth]{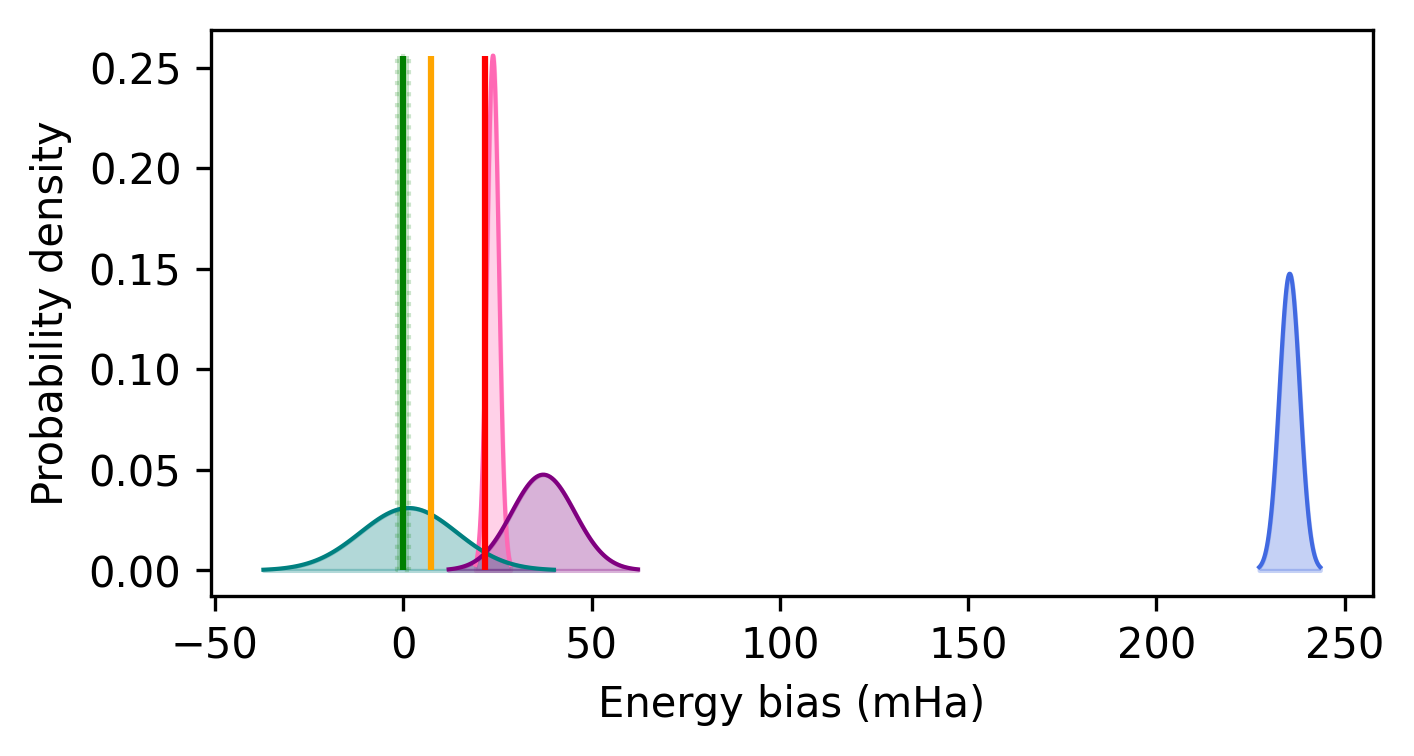}
    \caption{ibmq\_mumbai}\label{fig:mumbai_hist}
    \end{subfigure}
    \begin{subfigure}[t]{.45\linewidth}
    \includegraphics[width=\linewidth]{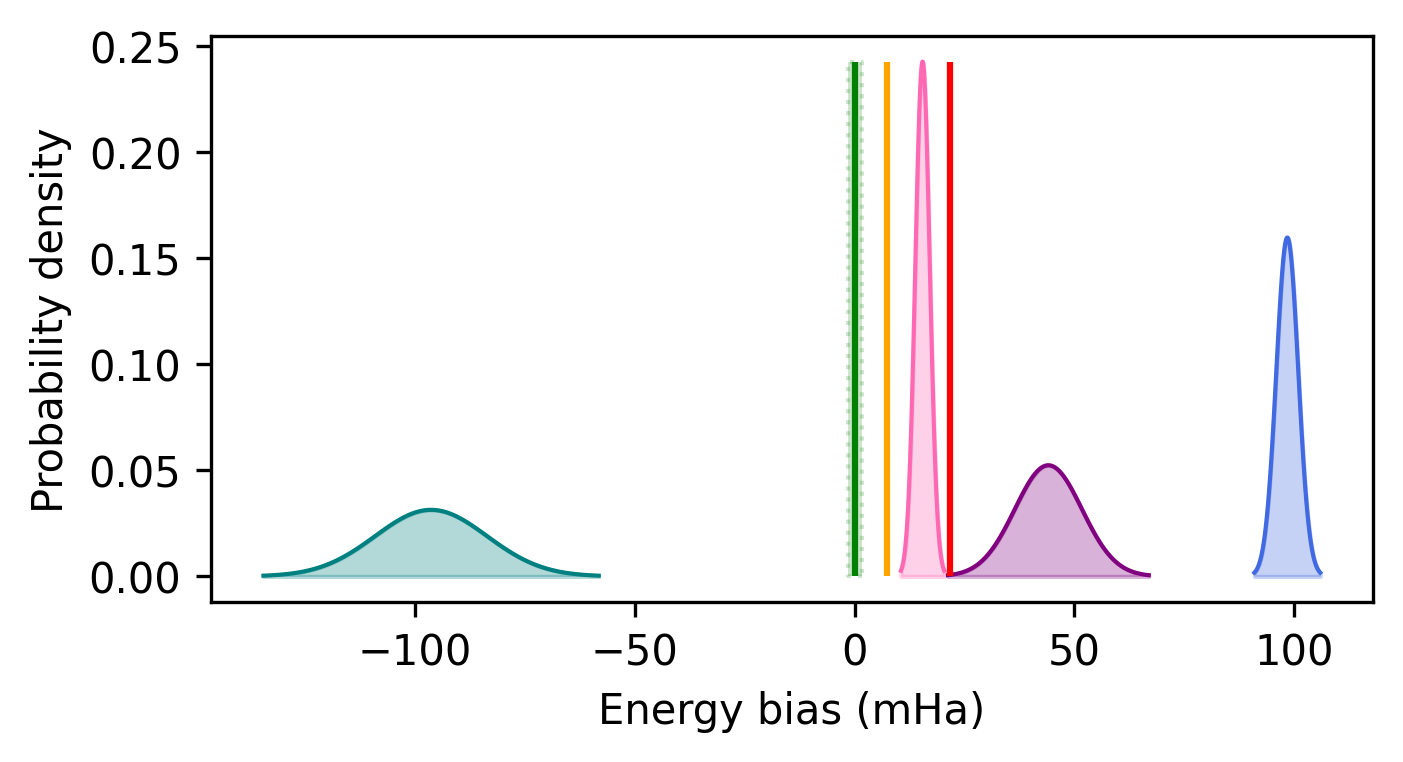}
    \caption{ibm\_hanoi}\label{fig:hanoi_hist}
    \end{subfigure}
    \begin{subfigure}[t]{.45\linewidth}
    \includegraphics[width=\linewidth]{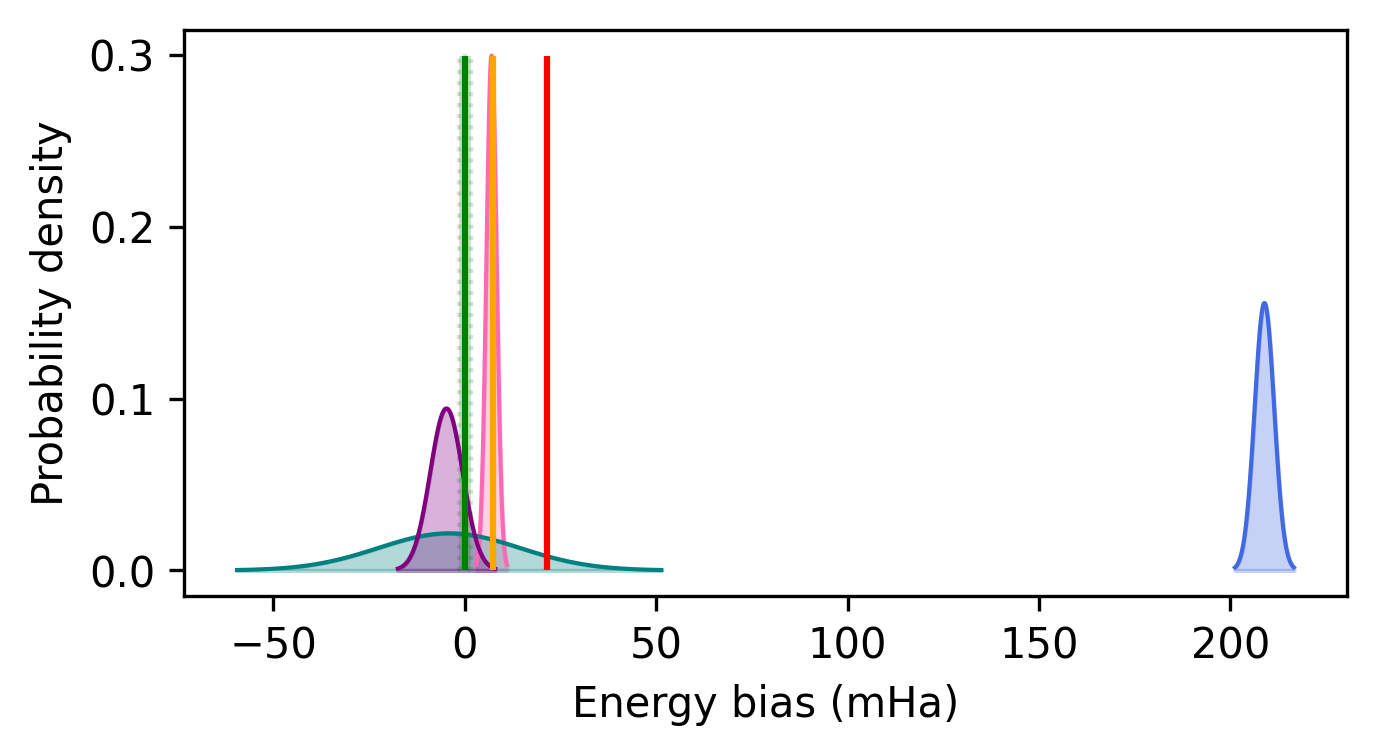}
    \caption{ibm\_cairo}\label{fig:cairo_hist}
    \end{subfigure}
    \begin{subfigure}[t]{.45\linewidth}
    \includegraphics[width=\linewidth]{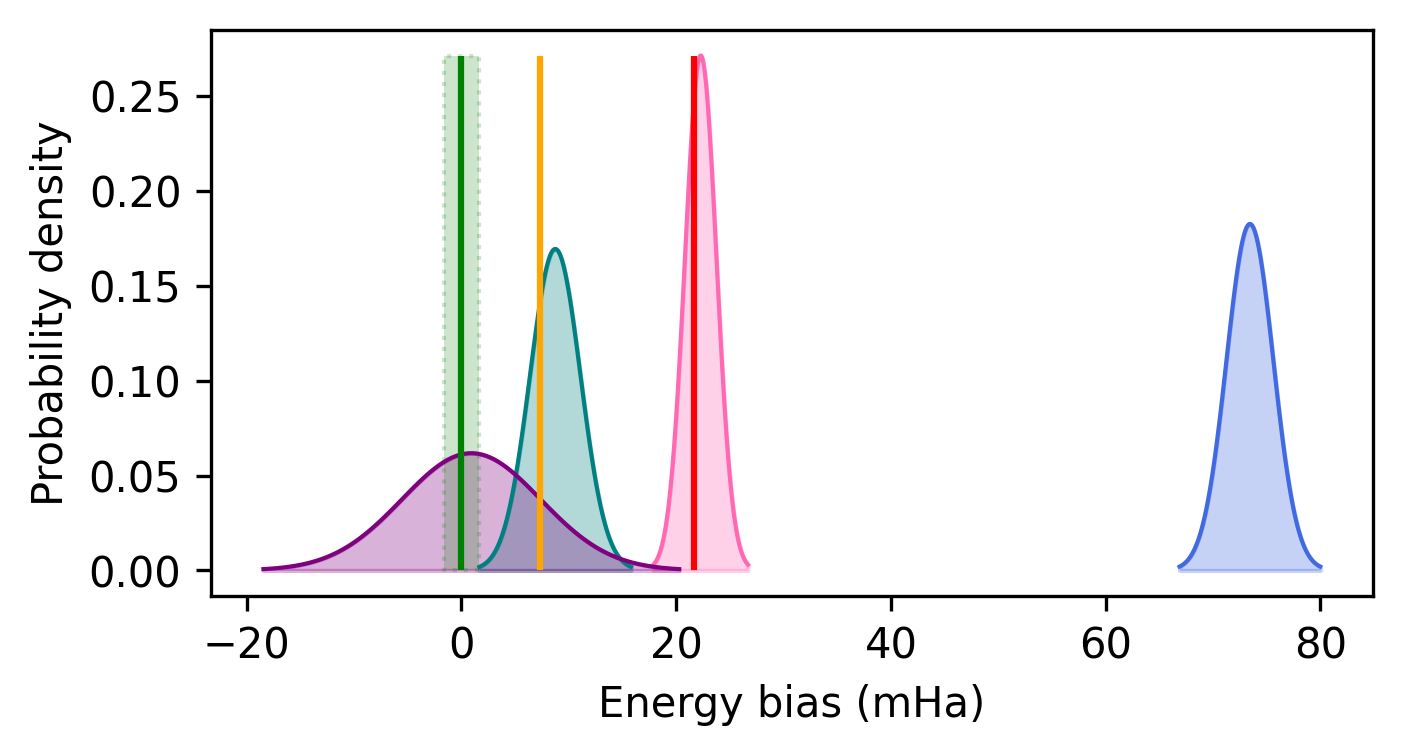}
    \caption{ibm\_auckland}\label{fig:auckland_hist}
    \end{subfigure}
    \begin{subfigure}[t]{.45\linewidth}
    \includegraphics[width=\linewidth]{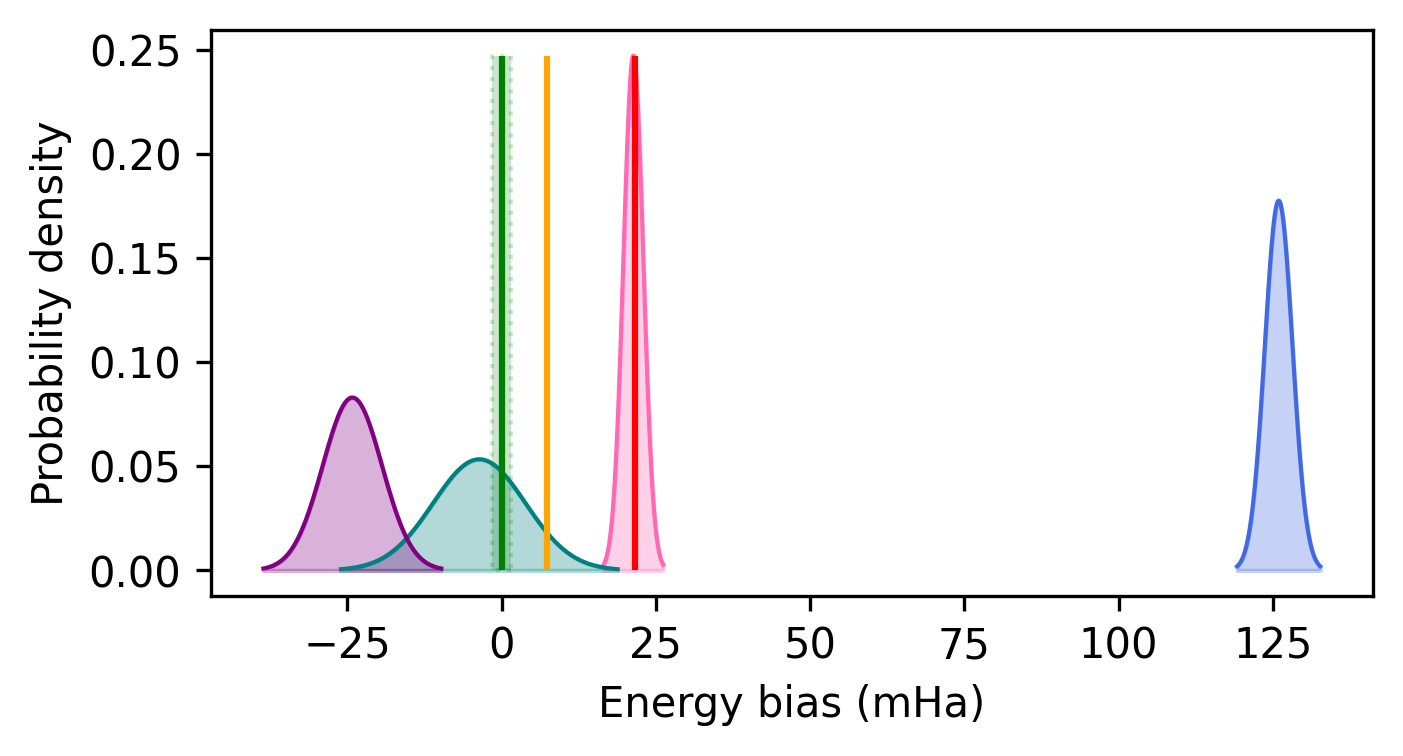}
    \caption{ibmq\_toronto}\label{fig:toronto_hist}
    \end{subfigure}
    \begin{subfigure}[t]{.45\linewidth}
    \includegraphics[width=\linewidth]{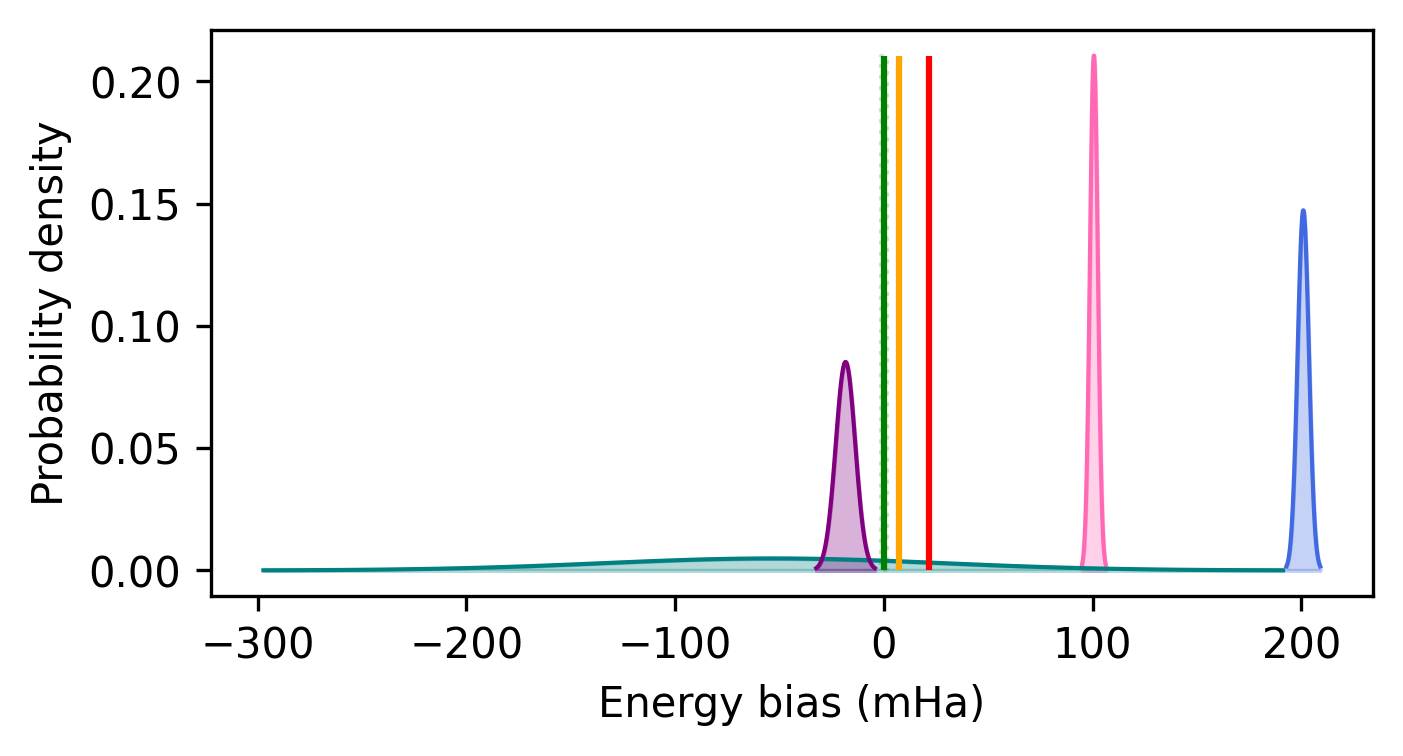}
    \caption{ibm\_geneva}\label{fig:geneva_hist}
    \end{subfigure}
    
    \begin{subfigure}[b]{\linewidth}
    \includegraphics[width=\linewidth]{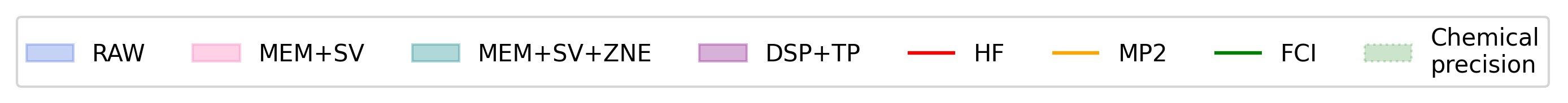}
    \end{subfigure}

\caption{Bootstrapped distributions for the best three QEM strategies identified through our benchmark. The mean energy of each distribution corresponds with the estimator bias. Note the failure of ZNE on \textit{ibm\_hanoi} and \textit{ibm\_geneva}, which is explained in Section \ref{sec:ZNE} and illustrates the sensitivity of this technique to erroneous fluctuations in the noise amplification.}
\label{fig:best_QEM_hist}

\end{figure*}

\begin{table*}
\centering
\caption{Comparison of estimator bias and standard deviation $\sigma$ (given in mHa) for various error-mitigation strategies performed across a suite of IBM Quantum 27-qubit Falcon devices with a shot budget $B=10^6$. The columns are ordered left-to-right by decreasing average error suppression (detailed in Table \ref{err_suppression}), with the exception of the raw estimate which is provided for reference. Note the following \textit{classical} quantum chemistry error benchmarks:\ HF - 21.621mHa, MP2 - 7.360mHa, Chemical Precision - 1.6mHa.}
\label{mitigation_benchmark}
\begin{tabular}{ll>{\columncolor[gray]{0.8}}cccccccccccccccc}
\toprule
            &          & \rotatebox{90}{RAW} & \rotatebox{90}{MEM+SV+ZNE} & \rotatebox{90}{DSP+TP} & \rotatebox{90}{MEM+DSP+TP} & \rotatebox{90}{MEM+ZNE} & \rotatebox{90}{MEM+SV} & \rotatebox{90}{SV+ZNE} & \rotatebox{90}{MEM} & \rotatebox{90}{SV} & \rotatebox{90}{MEM+DSP+TP+ZNE} & \rotatebox{90}{DSP+TP+ZNE} & \rotatebox{90}{ZNE} & \rotatebox{90}{MEM+DSP+ZNE} & \rotatebox{90}{MEM+DSP} & \rotatebox{90}{DSP+ZNE} & \rotatebox{90}{DSP} \\
\midrule
ibmq\_montreal & \textbf{bias} &      \textbf{294.7} &               \textbf{16.7} &           \textbf{0.8} &               \textbf{6.9} &           \textbf{19.0} &           \textbf{30.2} &           \textbf{14.9} &       \textbf{85.9} &       \textbf{59.2} &                  \textbf{61.1} &             \textbf{143.2} &      \textbf{240.7} &               \textbf{65.6} &           \textbf{72.4} &          \textbf{301.9} &      \textbf{208.1} \\
            & $\sigma$ &                 2.9 &                        13.4 &                    4.6 &                        3.7 &                    16.5 &                     1.7 &                    14.0 &                 2.0 &                 1.9 &                            9.0 &                       18.1 &                13.5 &                        10.8 &                     4.2 &                    16.8 &                 5.8 \\
ibmq\_kolkata & \textbf{bias} &       \textbf{85.5} &                \textbf{9.4} &           \textbf{1.4} &               \textbf{3.3} &            \textbf{1.8} &           \textbf{28.1} &           \textbf{40.3} &       \textbf{28.2} &       \textbf{57.1} &                   \textbf{8.8} &              \textbf{10.6} &       \textbf{65.5} &               \textbf{93.0} &           \textbf{37.3} &          \textbf{144.5} &       \textbf{88.7} \\
            & $\sigma$ &                 2.0 &                         2.2 &                    7.7 &                        6.9 &                     7.4 &                     1.5 &                     1.9 &                 1.5 &                 1.8 &                           17.5 &                       18.2 &                 4.7 &                        17.2 &                     6.9 &                    18.0 &                 7.4 \\
ibmq\_mumbai & \textbf{bias} &      \textbf{235.4} &                \textbf{1.4} &          \textbf{37.2} &               \textbf{3.3} &           \textbf{36.0} &           \textbf{23.8} &           \textbf{18.6} &       \textbf{42.9} &       \textbf{44.7} &                   \textbf{0.6} &              \textbf{54.5} &      \textbf{118.1} &              \textbf{146.7} &          \textbf{165.1} &          \textbf{332.4} &      \textbf{349.8} \\
            & $\sigma$ &                 2.7 &                        12.9 &                    8.4 &                        6.9 &                    12.9 &                     1.6 &                    15.5 &                 1.7 &                 1.7 &                           23.9 &                       25.7 &                12.7 &                        23.2 &                     6.7 &                    23.8 &                 7.7 \\
ibm\_auckland & \textbf{bias} &       \textbf{73.4} &                \textbf{8.7} &           \textbf{0.9} &               \textbf{7.1} &            \textbf{5.8} &           \textbf{22.3} &           \textbf{29.1} &       \textbf{25.4} &       \textbf{38.1} &                 \textbf{146.1} &             \textbf{152.6} &       \textbf{54.2} &              \textbf{248.8} &          \textbf{315.8} &          \textbf{293.0} &      \textbf{362.5} \\
            & $\sigma$ &                 2.2 &                         2.4 &                    6.5 &                        5.5 &                     2.6 &                     1.5 &                     0.8 &                 1.5 &                 1.7 &                           15.7 &                       15.8 &                 1.4 &                        16.4 &                     6.0 &                    16.5 &                 6.6 \\
ibm\_cairo & \textbf{bias} &      \textbf{208.9} &                \textbf{4.1} &           \textbf{4.8} &               \textbf{4.7} &           \textbf{64.4} &            \textbf{7.0} &           \textbf{26.5} &        \textbf{6.9} &       \textbf{64.0} &                  \textbf{19.5} &               \textbf{8.9} &       \textbf{99.8} &              \textbf{255.2} &          \textbf{200.4} &          \textbf{205.4} &      \textbf{242.8} \\
            & $\sigma$ &                 2.6 &                        18.5 &                    4.2 &                        5.3 &                    64.2 &                     1.3 &                    22.0 &                 1.3 &                 1.9 &                           20.8 &                       18.2 &                63.3 &                        19.6 &                     5.5 &                    17.5 &                 4.9 \\
ibm\_hanoi & \textbf{bias} &       \textbf{98.5} &               \textbf{96.4} &          \textbf{44.1} &              \textbf{31.1} &          \textbf{157.9} &           \textbf{15.5} &           \textbf{64.5} &       \textbf{20.8} &       \textbf{30.8} &                  \textbf{12.0} &              \textbf{26.6} &      \textbf{110.6} &               \textbf{59.6} &           \textbf{90.2} &           \textbf{84.7} &      \textbf{171.9} \\
            & $\sigma$ &                 2.5 &                        12.7 &                    7.6 &                        7.3 &                    23.8 &                     1.6 &                    12.2 &                 1.7 &                 1.9 &                           21.2 &                       20.1 &                23.4 &                        21.5 &                     7.2 &                    21.0 &                 7.2 \\
ibmq\_toronto & \textbf{bias} &      \textbf{125.9} &                \textbf{3.6} &          \textbf{24.2} &              \textbf{30.8} &           \textbf{18.8} &           \textbf{21.4} &           \textbf{11.0} &       \textbf{28.6} &       \textbf{37.9} &                   \textbf{2.3} &             \textbf{125.6} &       \textbf{87.7} &                \textbf{8.4} &           \textbf{55.0} &          \textbf{123.6} &      \textbf{162.4} \\
            & $\sigma$ &                 2.2 &                         7.5 &                    4.8 &                        1.8 &                     4.5 &                     1.6 &                    11.5 &                 1.7 &                 1.8 &                            9.9 &                       12.6 &                 7.2 &                        10.3 &                     2.4 &                    12.2 &                 5.0 \\
ibm\_geneva & \textbf{bias} &      \textbf{200.9} &               \textbf{53.0} &          \textbf{18.4} &               \textbf{3.6} &          \textbf{109.5} &          \textbf{100.6} &           \textbf{65.9} &      \textbf{100.6} &      \textbf{153.7} &                  \textbf{27.8} &              \textbf{37.9} &      \textbf{107.2} &              \textbf{358.7} &           \textbf{28.0} &          \textbf{471.2} &      \textbf{232.3} \\
            & $\sigma$ &                 2.7 &                        81.4 &                    4.7 &                        7.8 &                   148.4 &                     1.9 &                    24.3 &                 1.9 &                 2.3 &                           14.6 &                       12.3 &                60.4 &                        15.2 &                     7.8 &                    16.1 &                 5.5 \\
\bottomrule
\end{tabular}
\end{table*}

In Table \ref{mitigation_benchmark} we report the results of benchmarking our suite of error mitigation strategies for the 3-qubit \ce{HCl} problem across every 27-qubit system currently available to us through IBM Quantum with a shot budget of $B = 10^6$; the order in which each QEM technique (MEM, SV, ZNE, DSP, TP) appears in the combined strategy identifier indicates the order in which each method is being applied. Table \ref{err_suppression} presents the average error suppression in relation to the raw estimate, calculated as
\begin{equation}
    \bigg( 1 - \bigg| \frac{\stat{bias}{QEM}}{\stat{bias}{RAW}} \bigg| \bigg) \times 100 \%,
\end{equation} 
and change in standard deviation $\sigma$ across our suite of systems excluding \textit{ibm\_hanoi} and \textit{ibm\_geneva}, due to these systems performing sub-optimally (resulting in a failure of ZNE in Figure \ref{fig:ZNE_fits}). When $\stat{bias}{QEM}$ is near zero, the error suppression will approach $100\%$, whereas values close to $0\%$ indicate little (or no) improvement over the raw estimator; negative values of error suppression correspond with instances whereby the QEM strategy has had a detrimental effect to the energy estimate, a highly unfavourable situation.

The shot budget yields a raw standard deviation of $2 < \sigma < 3$ mHa, quantified via a bootstrapping procedure (discussed in Appendix \ref{sec:bootstrapping}). In Figure \ref{fig:best_QEM_hist} we plot the bootstrap distributions for a selection of the best performing QEM strategies to illustrate the trade-off between estimator bias and variance in practice, serving as a valuable comparison with previous theoretical analyses \cite{cai2022quantum}.

We observed that application of the MEM and SV techniques served to consistently lower both the estimator bias and standard deviation, which can be attributed to these approaches rectifying readout errors. Used in combination, the MEM+SV strategy permitted a respectable reduction in bias while also suppressing deviations with very little classical overhead. 

Unlike MEM and SV, the ZNE and DSP techniques necessitate modification to the quantum circuits themselves; the former, a decomposition of each CNOT gate into procedurally more complex circuit blocks, and the latter requiring a prepare-readout-invert structure with a supplementary ancilla qubit. Both of these methods can be seen to inflate the standard deviation.

By itself, DSP performs very poorly (indeed, the worst four strategies were all DSP-based), but when used in combination with TP we are permitted dramatic reductions in bias which exceed all other QEM strategies in the benchmark. The dependence on tomography purification for the ancilla qubit was also observed in Huo \& Li \cite{huo2022dual} and is essential to obtain good results from dual-state purification. We stress that, although state tomography is not scalable in general, here it is applied to a single qubit and hence does not contribute a significant cost in the number of measurements required. 

We found mixed success with ZNE-based strategies depending on which other QEM techniques were deployed in combination. Applied on top of MEM and SV we observed a significant improvement in error suppression, bar \textit{ibm\_hanoi} and \textit{ibm\_geneva} where extrapolation failed (Figure \ref{fig:ZNE_fits}), albeit at a significant increase in standard deviation. On the other hand, performing noise amplification on the ancilla qubit for the purposes of DSP produced disappointing results. These observations might be attributed to coherent errors causing unpredictable noise amplification behaviour; this could have been improved by including probabilistic error cancellation \cite{temme2017error}, converting coherent error into incoherent error that may be extrapolated more confidently.

\section{Conclusion}\label{sec:conclusion}

In this work we compared various quantum error-mitigation strategies, applied to the problem of preparing the \ce{HCl} molecule ground state on NISQ hardware. Motivated by the results of our benchmark in Section \ref{sec:results}, we identified three hybrid strategies with the strongest performance:
\begin{itemize}
    \item \textbf{Dual-State Purification with Tomography Purification (DSP + TP)} yields compelling error suppression ($93.253\%$ on average) although at an increase in standard deviation (2.543 times the raw value on average); given a generous shot budget and sufficient qubit connectivity this strategy should produce reliably accurate results. Implementing dual-state purification requires heavy modification to the ansatz resulting in doubled circuit depth, although the errors incurred here are suppressed. Further layering measurement-error mitigation produces a similar suppression in error although the increase in standard deviation is slightly less (2.113 times the raw value on average).
    \item \textbf{Measurement-Error Mitigation with Symmetry Verification (MEM + SV)} comes with very low overhead yet respectable error suppression ($82.678\%$ on average) on top of a reduction in standard deviation (0.638 times the raw value on average). Furthermore, there is no required modification to the ansatz circuit since both techniques operate solely on the binary measurement output. We recommend this strategy for restrictive shot budgets or where qubit topology does not permit the readout block needed for dual-state purification.
    \item \textbf{Zero-Noise Extrapolation on top of Measurement Error Mitigation with Symmetry Verification (MEM + SV + ZNE)} is sensitive to many factors but used carefully can yield excellent results ($94.327\%$ average error suppression when we exclude the cases where extrapolation failed in Figure \ref{fig:ZNE_fits}). There are many approaches to implementing ZNE, even extending to the pulse-level. On superconducting devices this might be preferable since it offers fine control over noise amplification. ZNE produced the largest inflation in standard deviation (3.680 times the raw value on average) and therefore a significantly greater shot budget would be necessary, due to error propagation in the extrapolation and since we evaluate several noise factors per expectation value.
\end{itemize}
As indicated by Table \ref{err_suppression}, each of these strategies achieved an average error suppression exceeding $80\%$ across the suite of 27-qubit IBM Quantum chips. Given the level of noise present on these devices, which is reflected in the raw energy estimates, the high bar of chemical precision would necessitate a suppression of $98.783\%$. This was obtained for three out of eight instances of DSP+TP (on the highest QV=128 systems \textit{ibmq\_montreal} and \textit{ibmq\_kolkata}, plus the QV=64 system \textit{ibm\_auckland}, with further device specifications given in Table \ref{hardware_specs}) and a single instance of MEM+SV+ZNE (on the QV=128 system \textit{ibmq\_mumbai}), bearing in mind that the standard deviation exceeds the chemically precise region and an increased shot budget would be necessary to counteract this. 

From the empirical results presented in this work, it is clear that we must rely heavily on methods of quantum error mitigation if we are to obtain usable results from NISQ hardware. Through our benchmark on the IBM Quantum 27-qubit Falcon processors, we have demonstrated the most effective combined strategies which we intend to take forward in our future quantum simulation work.

\section*{acknowledgements}
T.W. and A.R. acknowledge support from the Unitary Fund and the Engineering and Physical Sciences Research Council (EP/S021582/1 and EP/L015242/1, respectively). T.W. also acknowledges support from CBKSciCon Ltd., Atos, Intel and Zapata. W.K. and P.J.L. acknowledge  support  by the NSF STAQ project (PHY-1818914). W. K. acknowledges support from the National Science Foundation, Grant No. DGE-1842474. S.S. wishes to acknowledge financial support from the National Centre for HPC, Big Data and Quantum Computing” (Spoke 10, CN00000013). P.V.C. is grateful for funding from the European Commission for VECMA (800925) and EPSRC for SEAVEA (EP/W007711/1). Access to the IBM Quantum Computers was obtained through the IBM Quantum Hub at CERN with which the Italian Institute of Technology (IIT) is affiliated. We would also like to thank George Ralli, Andrew Tranter at Quantinuum, William Simon and Oliver Maupin at Tufts University, Marco Maronese at IIT, Michele Grossi and Oriel Kiss at CERN for valuable discussions during the development of this work.

\bibliography{main}

\appendix

\section{Bootstrapping}\label{sec:bootstrapping}

To evaluate the uncertainty in our energy estimates we rely on the statistical technique of bootstrapping \cite{efron1994introduction}. Ideally, one would perform quantum experiments many times to probe the `true' population, but from a practical standpoint this is not feasible due to the length of time required for each energy estimate (in our case $\approx 30$ minutes for a shot budget of $B=10^6$). Instead, we perform the experiment just once and generate resampled measurement data from the empirical distribution. This technique is widespread in statistics and makes the statistical analysis very convenient, not least as we may assume normality under the central limit theorem, which we verified using the \textit{normaltest} function in \textit{SciPy} \cite{2020SciPy-NMeth} that implements the D’Agostino-Pearson test \cite{d1973tests}.

Suppose we perform an $n$-shot quantum experiment and obtain a collection of binary measurement outcomes $M = \{m_1, \dots, m_n\}$ where $m_i \in \mathbb{Z}_2^N$. Our various QEM strategies combine these measurements in some way to yield an energy estimate $\E{}(M)$, but we would like to say something about the uncertainty in each estimator \textit{without} having to perform further experiments. The bootstrapping approach involves resampling from the empirical measurement distribution $M$, namely sampling elements $m^\prime_i \in M$ with replacement to form a new set of $n$ measurements $M^\prime$. We perform this process as many times as possible given the available compute resource, say $R \in \mathbb{N}$ repetitions, to approximate
\begin{equation}
    \stat{var}{} \approx \frac{1}{R^2} \sum_{r=1}^{R} \sum_{s>r}^{R} \Big( \E{}(M^\prime_r) - \E{}(M^\prime_s) \Big)^2;
\end{equation}
this is how we obtained the variances in Table \ref{mitigation_benchmark}.

\begin{figure}[b!]
    \centering
    \includegraphics[width=0.9\linewidth]{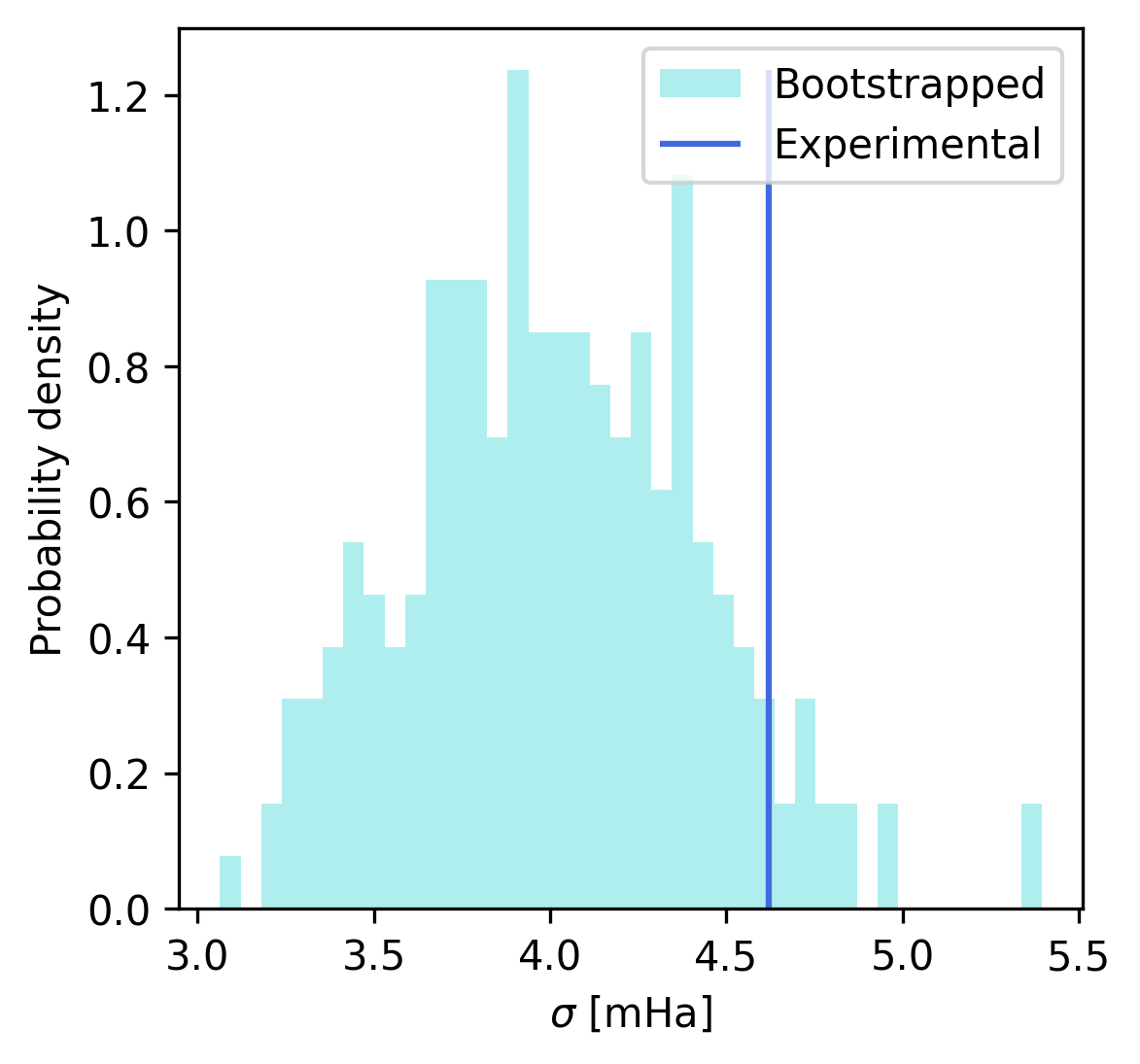}
    \caption{Distribution of bootstrapped standard deviations $\sigma$ versus the experimentally obtained value on \textit{ibmq\_kolkata}. We observe good agreement, with the bootstrapped values correct up to $\mathcal{O}(10^{-3}).$}
    \label{fig:BS_sigmas}
\end{figure}

One might question whether bootstrapping is well-motivated here. A priori, one has no reason to expect acceptable agreement with the true population parameters, hence we ran 225 instances of our quantum experiment applied just to the diagonal terms of the Hamiltonian (given in Table \ref{tab:3q_hamiltonian}), necessitating only computational basis measurements. We performed $10,000$ circuit shots in each experiment, for a combined total of $2.25 \times 10^6$ point samples before assessing the quality of the bootstrapped distributions against the overall sample. The 225 quantum experiments provide a \textit{target} standard deviation $\sigma$, indicated by the vertical line in Figure \ref{fig:BS_sigmas}, and we compare with this the \textit{bootstrap} standard deviations obtained per experiment.

\begin{figure}[b]
    \centering
    \includegraphics[width=0.9\linewidth]{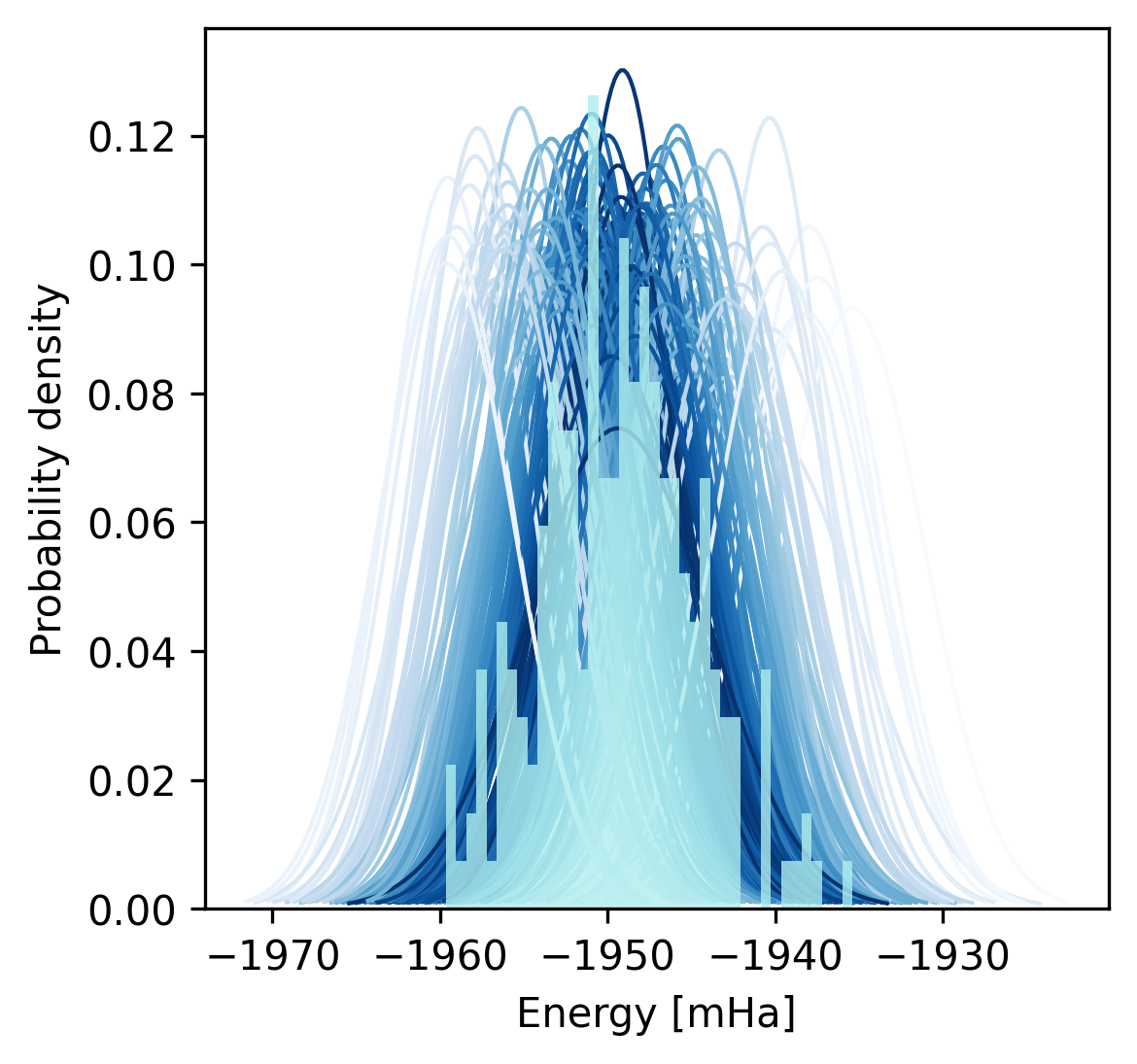}
    \caption{The true distribution of energy estimates obtained from 225 quantum experiments on \textit{ibmq\_kolkata}, each consisting of 10,000 circuit shots. Overlayed are the bootstrapped distributions for individual measurement sets to understand the relation between bootstrapping and the true population; the colour gradient indicates how far a given sample lies from the true (empirical) mean.}
    \label{fig:BS_test}
\end{figure}

In Figure \ref{fig:BS_test} we plot the result of our bootstrapping test and see reasonable agreement with the true energy distribution obtained from the NISQ hardware; the standard deviations all coincide with the experimentally-obtained value to $\mathcal{O}(10^{-3})$ (on the order of chemical precision), as indicated in Figure \ref{fig:BS_sigmas}, and therefore we employ bootstrapping with confidence.

\end{document}